\documentclass[12pt,aps,floats]{revtex4}

\begin{document}
\def\bigint{{\displaystyle\int}}
\def\simlt{\stackrel{<}{{}_\sim}}
\def\simgt{\stackrel{>}{{}_\sim}}

\title{
Solving the EPR paradox with pseudo-classical paths
}

\author{David H. Oaknin}
\affiliation{Rafael Ltd, IL-31021 Haifa, Israel \\
e-mail: d1306av@gmail.com \\
}

\begin{abstract}
We propose a novel interpretation of Quantum Mechanics, which can resolve the outstanding conflict between the principles of  locality and realism and offers new insight on the so-called weak values of physical observables. The discussion is presented in the context of Bohm's system of two photons in their singlet polarization state in which the Einstein-Podolski-Rosen paradox is commonly addressed. It is shown that quantum states can be understood as statistical mixtures of non-interfering pseudo-classical paths in a {\it hidden} phase space, in a way that overcomes the implicit assumptions of Bell's theorem and reproduces all expected values and correlations. The polarization properties of the photons along these paths are gauge-dependent magnitudes, whose actual values get fixed only after a reference direction is set by the observer of either photon A or B. Furthermore, these values are not constrained to fulfill standard classical algebraic relationships. These {\it hidden} paths can be grouped into coarser ones consistent with particular post-selection conditions for a complete set of commuting observables and along which every physical observable gets on average its corresponding weak value. Obviously, different sets of commuting observables lead to different coarse statistical representations of the same quantum state. This interpretation follows from the observation that  in the Heisenberg picture  of a closed quantum system in state $|\Psi>$ every physical observable ${\cal O}(t)=e^{+i H t} {\cal O} e^{-i H t}$  can be represented by an operator $P_{o(t)}$ within a commutative algebra, such that ${\cal O}(t)|\Psi>=P_{o(t)}|\Psi>$.  The formalism presented here may become a useful tool for performing numerical simulations of quantum systems.
\end{abstract}

\maketitle
{\bf 1.}  Quantum Mechanics is widely believed to be the ultimate theoretical framework within which all fundamental laws of Nature are to be formulated. Its postulates have been extensively and accurately tested in a very broad class of physical systems, including  optics, atomic and molecular physics, condensed matter physics and high-energy particle physics. Indeed, one of the greatest challenges in physics at present is to formulate Einstein's general relativity theory of gravitation within this framework. 

Nevertheless, there remain crucial questions about the interpretation of Quantum Mechanics that have not been properly understood yet. In particular, it has been known since long ago that the current interpretation of the quantum formalism cannot accommodate together two fundamental principles of modern physics usually taken for granted, namely, the principle of locality and the principle of physical realism. The principle of locality states that physical events cannot affect or be affected by other events in space-like separated regions. Besides the principle of realism claims that all measurable physical observables correspond to intrinsic properties of the physical world, which a final theory should be able to completely account for.  The clash between these two principles was first noticed by Einstein, Podolsky and Rosen \cite{EPR} and it has been known since then as the EPR paradox. The paradox is commonly formulated as follows \cite{Bohm}:

Consider a massive particle that decays into two distinguishable photons, which travel in opposite directions along the Z axis. The two photons are assumed to be emitted in their singlet polarization state:

\begin{equation}
\label{Bell_state}
| \Psi > = \frac{1}{\sqrt{2}} \left(| \uparrow \ \ \downarrow > - | \downarrow \ \ \uparrow > \right),
\end{equation} 
where $\{| \uparrow >, | \downarrow > \}$ is an orthonormal linear basis in the single particle polarization Hilbert space.  
If the photons travel freely, their polarization state remains entangled once they have travelled far away from each other.

In this state the polarizations of the photons are perfectly anti-correlated when they are measured along parallel directions: if we would perform a measurement of the polarization of photon A, we would know with certainty also the polarization of photon B along that direction. Hence, accepting the principle of locality implies that we can gain certainty on the polarization of photon B along any direction without perturbing it in any sense. Furthermore, accepting the principle of physical realism implies then that the polarization properties of photon B were set at emission. Obviously, the same could be said about the polarization of photon A. Nevertheless, according to the current interpretation of Quantum Mechanics nothing can be said with certainty about the polarization of each of the photons from their wavefunction at emission (\ref{Bell_state}). This observation lead Einstein, Podolsky and Rosen \cite{EPR} to claim that the description of the physical system provided by the wavefunction is not complete. 

The description of the quantum system would be complete if we could interpret the wavefunction as a statistical mixture of classical paths defined in some more fundamental {\it hidden variables} phase space, so that a measurement of the polarization of any of the two photons would imply nothing but a mere update of our knowledge about the state of the system. Unfortunately, Bell's theorem \cite{Bell} rules out the possibility to build such a statistical classical interpretation of the wavefunction based on the currently accepted notions of physical realism and locality. Indeed, the theorem proves that in any model of classical {\it hidden variables} based on these premises there exist certain constraints on the statistical correlations between physical observables (Bell's inequalities), which are nonetheless not necessarily fulfilled by their quantum mechanical counterparts. Additional inequalities of this kind  (CHSH inequalities) discovered later on by Clauser, Horne, Shimony and Holt \cite{CHSH} and Clauser and Horne \cite{CH} actually allowed to experimentally verify the predictions of quantum mechanics against these statistical models of classical local {\it hidden variables}.

Formally, Bell's theorem can be stated as follows. Let us assume that the two photons system could be described as a statistical mixture of classical paths with well defined probabilities $\rho(\lambda)$, where the label $\lambda \in {\cal S}$ sets the final conditions of the paths in the {\it hidden variables} phase space ${\cal S}$. Along each path every physical observable $O$ has a well defined value $o(\lambda)$, which is assumed to belong to its spectrum of eigenvalues. In particular, the spin polarization of photon A along any direction ${\vec a}$ in the unit sphere - denoted as $\sigma^{(A)}({\vec a}, \lambda)$  -  is assumed to get values either $+1$ or $-1$ on each of the paths. Similarly, the spin polarization of photon B along any other direction ${\vec b}$  - denoted as $\sigma^{(B)}({\vec b}, \lambda)$. The stated constraint that requires the photons polarizations to be perfectly anti-correlated when measured along the same direction demands $\sigma^{(B)}({\vec a}, \lambda) = - \sigma^{(A)}({\vec a},\lambda)$.  Therefore, the expected correlation between the spin polarizations of the two photons when measured along any two arbitrary directions is given by:

\begin{equation}
E({\vec a}, {\vec b}) \equiv \int \ d\lambda \ \rho(\lambda)  \ \sigma^{(A)}({\vec a},\lambda) \ \sigma^{(B)}({\vec b},\lambda) = - \int \ d\lambda \ \rho(\lambda)  \ \sigma^{(A)}({\vec a},\lambda) \ \sigma^{(A)}({\vec b},\lambda). 
\end{equation}
Bell's theorem states that  for any triplet of unit vectors ${\vec a}$,  ${\vec b}$ and ${\vec c}$ the following inequality holds:

\begin{equation}
\label{Bell_theorem}
\left| E({\vec a}, {\vec b}) - E({\vec a}, {\vec c}) \right| \le 1 + E({\vec b}, {\vec c}).
\end{equation}

The importance of Bell's statement (\ref{Bell_theorem}) lies on the fact that the correlations between the spin polarizations of the two photons predicted by Quantum Mechanics $E({\vec a}, {\vec b}) = <\Psi|\sigma^{(A)}_{\vec a}\cdot\sigma^{(B)}_{\vec b}|\Psi>=- {\vec a} \cdot {\vec b}$ are not constrained by this inequality. For example, for ${\vec b} = {\vec e}_x = (1,0,0)$,  ${\vec c} = {\vec e}_y = (0,1,0)$ and ${\vec a} = \frac{1}{\sqrt{2}}(1,-1,0)$, we find that,  $\left| E({\vec a}, {\vec b}) - E({\vec a}, {\vec c}) \right| = \left|{\vec a} \cdot \left({\vec c} - {\vec b}\right) \right| = \sqrt{2}$, while $ 1 + E({\vec b}, {\vec c}) = 1$. Hence, the statistical models of classical {\it hidden} variables considered by Bell cannot completely reproduce the predictions of Quantum Mechanics. 

In the past, most of the efforts to solve the issue of the completeness of the quantum mechanical description of the physical world focused on exploring the consequences of giving up in some way the principle of locality. Nevertheless, it has been shown that neither a general class of appealing non-local theories of classical {\it hidden variables} \cite{Leggett} can reproduce all quantum correlations \cite{Zeilinger}. The proof of this statement is based on a generalized version of the Clauser-Horne-Shimony-Holt inequality \cite{CHSH} that holds within all these non-local {\it hidden variables} models, but it is not necessarily fulfilled by quantum mechanics. These results have led some leading physicists \cite{Zeilinger} to suggest that it might be necessary to abandon certain intuitive aspects of the currently accepted notion of physical realism. 

In this paper we demonstrate that the difficulties to integrate together within the framework of Quantum Mechanics both the principle of locality and the principle of physical realism are a consequence of some implicit assumptions about the latter. These difficulties are removed once these assumptions are lifted. Namely, most of the attempts made to date to interpret the quantum wavefunction as a statistical mixture of non-interfering paths have implicitly assumed that the actual values of the polarization properties of the photons in these paths can be defined independently of the direction along which they are observed and must be equal to one of the eigenvalues of the given observable and, therefore, obey standard classical algebraic relationships. Indeed, the proof of Bell's inequality  (\ref{Bell_theorem}) crucially relies on these assumptions, as $\sigma^{(A)}({\vec k},\lambda)$ is constrained to take values either $+1$ or $-1$ along any direction ${\vec k}$ and, therefore, $\sigma^{(A)}({\vec b},\lambda) \cdot \sigma^{(A)}({\vec b},\lambda) = 1$ and   $\left| \sigma^{(A)}({\vec b},\lambda) \cdot \sigma^{(A)}({\vec c},\lambda) \right| = 1$. Hence,

\begin{eqnarray*}
\label{Bell_inequality}
\left| E({\vec a}, {\vec b}) - E({\vec a}, {\vec c}) \right| & = & \left| \int \ d\lambda \ \rho(\lambda)  \left[ \sigma^{(A)}({\vec a},\lambda) \ \sigma^{(A)}({\vec b},\lambda) - \sigma^{(A)}({\vec a},\lambda) \ \sigma^{(A)}({\vec c},\lambda) \right] \right|= \\
& =  & \left| \int \ d\lambda \ \rho(\lambda)  \left[ \sigma^{(A)}({\vec a},\lambda) \ \sigma^{(A)}({\vec b},\lambda) - \sigma^{(A)}({\vec a},\lambda) \ \sigma^{(A)}({\vec b},\lambda) \ \sigma^{(A)}({\vec b},\lambda) \ \sigma^{(A)}({\vec c},\lambda) \right] \right|= \\
& =  & \left| \int \ d\lambda \ \rho(\lambda) \ \sigma^{(A)}({\vec a},\lambda) \ \sigma^{(A)}({\vec b},\lambda) \left[1 - \sigma^{(A)}({\vec b},\lambda) \ \sigma^{(A)}({\vec c},\lambda) \right] \right| \\ & \le & \int \ d\lambda \ \rho(\lambda) \  \left(1 - \sigma^{(A)}({\vec b},\lambda) \ \sigma^{(A)}({\vec c},\lambda) \right) = 1 + E({\vec b}, {\vec c}).
\end{eqnarray*}

These implicit assumptions may not be justified. They stem from the Von Neumann paradigm of strong (projective) measurements, whose only possible outcomes are any of the eigenvalues of the measured observable. 
Notwithstanding, Bell's theorem requires to estimate the polarization of photon A along three distinct directions, ${\vec a}$, ${\vec b}$ and ${\vec c}$, while we can actually perform  strong measurements of its polarization along at most two directions, say ${\vec a}$ and ${\vec b}$: one component is obtained by directly measuring on this photon and the second component is obtained by measuring on the second photon and exploiting the perfect anti-correlation between the two. The only experimental access that we can have to the photon polarization along a third direction ${\vec c}$ is through weak measurements, whose output  can have absolute values larger and smaller than one and may even be complex \cite{Jozsa}. Moreover, the weak value of the polarization of the photon along this third direction depends on the choice of the two directions along which its polarization components are  strongly measured. Such weak values of the polarization of single photons have been experimentally measured and confronted with theoretical predictions in different setups \cite{Parks, Pryde}. 

By giving up these implicit assumptions we generalize the notion of physical realism and avoid the constraints set up by Bell's theorem \cite{Fine}. It is then fairly easy to give to the quantum wavefunction (\ref{Bell_state}) a statistical interpretation that reproduces the average values and correlations of all physical observables. We proceed as follows: Let denote by $|\zeta_{\pm,\pm} >$ the quantum eigenstates for a complete set of commuting observables $\{\sigma^{(A)}_{\vec a}$, $\sigma^{(B)}_{\vec b}\}$, where the labels $\pm,\pm$ indicate their corresponding eigenvalues. The singlet Bell state (\ref{Bell_state}) can be described as a statistical mixture of four pseudo-classical paths, which we will label as $\lambda=++, +-,-+,--$, such that:

\begin{itemize}
\item  Each one of of the four paths happens with probability 

\begin{equation}
\label{Prob_path}
\rho(\pm,\pm)=|<\zeta_{\pm,\pm}|\Psi>|^2,
\end{equation}
so that, $\sum_{\lambda=\pm,\pm} \rho(\lambda) = 1$. 

\item Along each of these paths every physical observable ${\cal O}(t)=e^{+i H t} {\cal O} e^{-i H t}$ in the Heisenberg picture is given its corresponding weak value  \cite{AAV, AV, AACV, aharonov, Duck}:

\begin{equation}
\label{weak_value}
O_w(t)(\pm,\pm) = \frac{<\zeta_{\pm,\pm}|{\cal O}(t)|\Psi>}{<\zeta_{\pm,\pm}|\Psi>}.
\end{equation}
\end{itemize}

It is straightforward to show that this statistical model reproduces the quantum average value of any physical observable: 

\begin{eqnarray*}
\sum_{\pm,\pm} \rho(\pm,\pm) \cdot  {\cal O}_{\it w}(t)({\pm,\pm}) = \sum_{\pm,\pm} |<\zeta_{\pm,\pm}|\Psi>|^2 \ \frac{<\zeta_{\pm,\pm}|{\cal O}(t)|\Psi >}{< \zeta_{\pm,\pm}|\Psi >}= \\
=\sum_{\pm,\pm} <\Psi|\zeta_{\pm,\pm}> <\zeta_{\pm,\pm}|{\cal O}(t)|\Psi >= <\Psi|{\cal O}(t)|\Psi>.
\end{eqnarray*} 
and also the quntum correlations between any two physical observables ${\cal O}_1(t_1)$ and ${\cal O}_2(t_2)$: 

\begin{eqnarray*}
\sum_{\pm,\pm}   \rho(\pm,\pm)  \  ({\cal O}_1)_w(t_1)^*({\pm,\pm}) ({\cal O}_2)_w(t_2)({\pm,\pm})=  \hspace{2.0in}\\
=\sum_{\pm,\pm} |<\zeta_{\pm,\pm}|\Psi>|^2  \left(\frac{<\zeta_{\pm,\pm}|{\cal O}_1(t_1)|\Psi >}{< \zeta_{\pm,\pm}|\Psi >}\right)^* \left(\frac{<\zeta_{\pm,\pm}|{\cal O}_2(t_2)|\Psi >}{< \zeta_{\pm,\pm}|\Psi >}\right)= \\
=\sum_{\pm,\pm} <\Psi|{\cal O}_1(t_1)|\zeta_{\pm,\pm}> <\zeta_{\pm,\pm}|{\cal O}_2(t_2)|\Psi >= <\Psi|{\cal O}_1(t_1) \cdot {\cal O}_2(t_2)|\Psi>.
\end{eqnarray*}
Furthermore, this statistical interpretation of the quantum state is explicitly local: the values (\ref{weak_value}) along these paths of observable properties of one of the photons, say photon $A$, do not change when photon $B$ interacts with the external world. This statement can be easily proved by noticing that the hermitic operators ${\cal O}_A$ that describe physical properties of photon $A$ do commute with the hamiltonian $H_B$ that describes the interaction of photon $B$ with the external world and, therefore, ${\cal O}_A(t)=e^{+i H_B t} {\cal O}_A e^{-i H_B t} = {\cal O}_A$. Hence,  $\left({\cal O}_A\right)_w(t)(\pm,\pm) = \left({\cal O}_A\right)_w(\pm,\pm)$.

On the other hand, we realize that each complete set of commuting observables leads to a different statistical representation (\ref{Prob_path}, \ref{weak_value}) of the same quantum state (\ref{Bell_state}) and we need to ask if all these possible representations can describe a unique common underlying physical reality. This question is answered in sections 4 and 5, where we show that the statistical representations associated to different complete sets of commuting observables $\{\sigma^{(A)}_{\vec a}, \sigma^{(B)}_{\vec b}\}$ that share a common operator, say $\sigma^{(A)}_{\vec a}$, are indeed different coarse descriptions of a common finely resolved statistical representation of the quantum state. That is, the pseudo-classical paths associated to different complete sets of commuting observables are coarse descriptions of finely resolved paths in a {\it hidden} phase space, consistent with particular post-selection conditions for the specified observables. The need to choose a common operator  $\sigma^{(A)}_{\vec a}$ to describe this {\it hidden} reality can be easily understood as a gauge-fixing condition for each particular observer, implying that the polarization properties of the photons in the finely resolved {\it hidden} phase space are indeed gauge-dependent magnitudes.  We want to stress, nonetheless, that our coarse pseudo-classical paths are essentially different from the coarse paths envisioned in the {\it consistent histories} interpretation of Quantum Mechanics \cite{Griffiths1,Griffiths2,Griffiths3,Omnes1,Omnes2,Gell-Mann,Halliwell}, in the sense that along each pseudo-classical path every quantum observable ${\cal O}(t)$ has a well defined value (\ref{weak_value}). Furthermore, in the {\it consistent histories} interpretation the time evolution of physical observables is intrinsically assumed to be stochastic \cite{Griffiths_new}, while along our pseudo-classical paths it is deterministic. In this sense, our interpretation of Quantum Mechanics in terms of pseudo-classical paths is closer to the approach advocated in \cite{t'hooft}.

Our statistical interpretation of the quantum wavefunction $|\Psi>$ can be easily understood by noticing (see section 2) that in the Heisenberg picture every physical observable ${\cal O}(t)=e^{+i H t} {\cal O} e^{-i H t}$  can be represented  by an operator $P_{o(t)}$ within a commutative algebra, such that \cite{david1,david2}

\begin{equation}
\label{the_equivalence}
{\cal O}(t)|\Psi>=P_{o(t)}|\Psi>. 
\end{equation}
    
Hence, once we choose an orthonormal basis of eigenstates of this algebra we can associate  to every observable well defined values. In fact, the eigenvalues of the operator $P_{o(t)}$ are equal to the weak values  (\ref{weak_value}) of its corresponding observable ${\cal O}(t)$ when the system is post-selected to each of these eigenstates.

As we noticed above the commutative algebra that represents the quantum system in this picture is not unique. Indeed, it is closely related to the complete set of commuting observables chosen to strongly measure the quantum system. Different choices of this set lead to equivalent commutative algebras to represent the system according to (\ref{the_equivalence}).  As we have mentioned above, we shall show that the algebras associated to complete sets of commuting observables that share one of its operators can be understood as coarse representations of a unique finely resolved algebra common to all of them. It is important to stress, nonetheless, that these algebras, in spite of being commutative, are not classical agebras, because for any pair of observables, ${\cal O}_1(t)$  and ${\cal O}_2(t)$, the commutative operator representing their product, either $P_{\{o_1(t), o_2(t)\}/2}$ or $P_{i [o_1(t), o_2(t)]/2}$,  is not necessarily equal to the product of the operators representing each one of them, $\{P_{o_1(t)}^{\dagger}, P_{o_2(t)}\}/2$ or $i [P_{o_1(t)}^{\dagger}, P_{o_2(t)}]/2$. In section 7 we shall discuss the conditions under which these equalities are recovered. These conditions can thus be understood as the onset of classicality, without any reference to external observers or environment.

The paper is organized as follows. In sections 2 and 3 we build the formalism of pseudo-classical paths for Bohm's  system of two entangled photons in their singlet polarization state. In sections 4 and 5 we show how these paths can be interpreted as coarse descriptions of a finely resolved statistical representation of the quantum state. 
In section 6 we discuss within this formalism the EPR paradox and the problem of the collapse of the wavefunction due to measurement. In section 7 we briefly explore the onset of classicality. In section 8 we summarize our conclussions.   Section 9 is an appendix where we briefly review the notions of strong and weak quantum measurements.
\\

{\bf 2.} We consider a system of two distinguishable photons travelling in opposite directions and whose polarizations are described by the singlet Bell state (\ref{Bell_state}). The axis along which the photons travel is labelled without any loss of generality as $Z$ axis.

We start our programme by choosing a complete set of commuting observables on the Hilbert space ${\cal H} \equiv {\cal H}_A \otimes {\cal H}_B$ of two photons polarization states. Such observables can be simultaneously measured through strong (projective) measurements and their common eigenstates set up an orthonormal basis in the Hilbert space.  Let us pick, for example, the polarization components of each of the two photons along two arbitrary directions in the plane $XY$ orthogonal to the direction they are moving along:  

\begin{equation}
\label{CSCO}
\{\sigma_1^{(A)}, \ \ \ \ \sigma_{\phi}^{(B)} \equiv \cos\left(\phi\right) \ \sigma_1^{(B)} + \sin\left(\phi\right) \sigma_2^{(B)}\},
\end{equation}
with $\phi \in \left(0, \pi\right) \bigcup \left(\pi, 2\pi \right)$ and where 

\[ 
\sigma_1 = \left( \begin{array}{cc}
0 & 1 \\
1 & 0
\end{array}
\right),
 \\  
\sigma_2 = \left( \begin{array}{cc}
0 & -i \\
i & 0
\end{array}
\right),
 \\  
\sigma_3 = \left( \begin{array}{cc}
1 & 0 \\
0 & -1
\end{array}
\right) \]
are Pauli matrices acting on single particle polarization states defined in the orthogonal basis $\{| \uparrow > \equiv \left( \begin{array}{c} 1 \\ 0 \end{array}\right), | \downarrow > \equiv \left( \begin{array}{c} 0 \\ 1 \end{array}\right)\}$ and the upper index $(A,B)$ indicates on which of the two photons the observable is defined. Without any loss of generality we have labelled the direction along which the polarization of photon A is strongly defined as $X$ axis.

We now  obtain the orthonormal basis of common eigenstates to the chosen complete set of commuting operators (\ref{CSCO}):

\begin{equation}
\label{E1}
|+ ; +> = \frac{1}{\sqrt{2}} \left(| \uparrow > + | \downarrow >\right)_A \otimes \frac{1}{\sqrt{2}} \left(| \uparrow > + e^{i \phi} | \downarrow >\right)_B
\end{equation}
\begin{equation}
\label{E2}
|+ ; -> = \frac{1}{\sqrt{2}} \left(| \uparrow > + | \downarrow >\right)_A \otimes \frac{1}{\sqrt{2}} \left(| \uparrow > - e^{i \phi} | \downarrow >\right)_B
\end{equation}
\begin{equation}
\label{E3}
|- ; +> = \frac{1}{\sqrt{2}} \left(| \uparrow > - | \downarrow >\right)_A \otimes \frac{1}{\sqrt{2}} \left(| \uparrow > + e^{i \phi} | \downarrow >\right)_B
\end{equation}
\begin{equation}
\label{E4}
|- ; -> = \frac{1}{\sqrt{2}} \left(| \uparrow > - | \downarrow >\right)_A \otimes \frac{1}{\sqrt{2}} \left(| \uparrow > - e^{i \phi} | \downarrow >\right)_B
\end{equation}
Each one of the two observables in the complete set that we have chosen has eigenvalues $\pm 1$. The notation chosen for their common eigenvectors indicates their eigenvalues for each one of the two observables: the first sign refers to operator $\sigma_1^{(A)}$ and the second to operator $\sigma_{\phi}^{(B)}$.

The singlet Bell state (\ref{Bell_state}) is now written in this basis:

\begin{equation}
\label{Bell_state_2}
| \Psi >  = q_{++} |+;+> + q_{+-} |+;-> + q_{-+} |-;+> + q_{--} |-;->,
\end{equation}
where 
\begin{eqnarray*}
q_{++} = <+;+|\Psi> =-\frac{1}{2 \sqrt{2}} \left(1 - e^{- i \phi}\right), \hspace{0.5in}
q_{+-} = <+;-|\Psi> =-\frac{1}{2 \sqrt{2}} \left(1 + e^{- i \phi}\right), \\
q_{-+} = <-;+|\Psi> =+\frac{1}{2 \sqrt{2}} \left(1 + e^{- i \phi}\right), \hspace{0.5in}
q_{--} = <-;-|\Psi> =+\frac{1}{2 \sqrt{2}} \left(1 - e^{- i \phi}\right), \\
\end{eqnarray*}
such that, 
\begin{eqnarray}
\label{probabilities_123}
\left| q_{++} \right|^2 = \left| q_{--} \right|^2 = \frac{1}{4} \left(1 - \cos \left(\phi\right)\right), \hspace{0.5in}
\left| q_{+-} \right|^2 = \left| q_{-+} \right|^2 = \frac{1}{4} \left(1 + \cos \left(\phi\right)\right).
\end{eqnarray}

The next and crucial step is to show how the wavefunction (\ref{Bell_state_2}) can be understood as a statistical mixture  of four non-interfering paths, denoted as $++$, $+-$, $-+$, $--$, each one occurring with well defined probability $|q_{\pm \pm}|^2$. In order to do it we will lay down a well defined set of rules that will allow us to assign to each physical observable a time-dependent c-value on each one of these paths. We will refer to it as the pseudo-classical value of the physical observable along the path.

To physical observables that commute with $\sigma_1^{(A)}$, $\sigma_{\phi}^{(B)}$ we obviously assign their eigenvalues for each one of the four eigenvectors. For example,

\begin{eqnarray}
\label{assignments1}
\left(\sigma_1^{(A)}\right)_{cl}(+, \pm) = +1, \ \ \ \ \ \ 
\left(\sigma_1^{(A)}\right)_{cl}(-, \pm) = -1, \\
\label{assignments1b}
\left(\sigma_{\phi}^{(B)}\right)_{cl}(\pm, +) = +1, \ \ \ \ \ \ 
\left(\sigma_{\phi}^{(B)}\right)_{cl}(\pm, -) = -1,
\end{eqnarray}
and also,
\begin{equation}
\label{assignments2}
\left({\bf 1}\right)_{cl}(\pm, \pm) = +1,
\end{equation}
\begin{equation}
\label{assignents3}
\left(\sigma_1^{(A)} \cdot \sigma_{\phi}^{(B)}\right)_{cl}(\pm, \pm) = \left(\sigma_1^{(A)}\right)_{cl}(\pm, \pm) \cdot
\left(\sigma_{\phi}^{(B)}\right)_{cl}(\pm, \pm).
\end{equation}

We now need to define the way to assign pseudo-classical values on paths to observables that do not commute with $\sigma_1^{(A)}$, $\sigma_{\phi}^{(B)}$. To do it we notice that

\begin{eqnarray*}
{\bf 1}|\Psi > & = & \frac{1}{\sqrt{2}} \left( |\uparrow \ \downarrow > - |\downarrow \ \uparrow >\right),  
\\
\sigma_1^{(A)}|\Psi > & = & \frac{1}{\sqrt{2}} \left( |\downarrow \ \downarrow > - |\uparrow \ \uparrow >\right), 
\\
\sigma_{\phi}^{(B)}|\Psi > & = & \frac{1}{\sqrt{2}} \left(e^{- i \phi} |\uparrow \ \uparrow > - e^{+ i \phi} |\downarrow \ \downarrow >\right),
\\  
\sigma_1^{(A)} \cdot \sigma_{\phi}^{(B)}|\Psi > & = & \frac{1}{\sqrt{2}} \left(e^{- i \phi} |\downarrow \ \uparrow > - e^{+ i \phi} |\uparrow \ \downarrow >\right),
\end{eqnarray*}
are four linearly independent vectors, whenever $\phi \in \left(0, \pi\right) \bigcup \left(\pi, 2\pi \right)$. Therefore, for any given observable ${\cal O}$ there exists one and only one linear combination ${\cal P}_o\left({\bf 1}, \sigma_1^{(A)}, \sigma_{\phi}^{(B)}, \sigma_1^{(A)} \cdot \sigma_{\phi}^{(B)}\right)$ of the four commuting operators ${\bf 1}$, $\sigma_1^{(A)}$, $\sigma_{\phi}^{(B)}$ and $\sigma_1^{(A)} \cdot \sigma_{\phi}^{(B)}$, such that, 

\begin{eqnarray}
\label{the_law}
{\cal O}|\Psi > = {\cal P}_o\left({\bf 1}, \sigma_1^{(A)}, \sigma_{\phi}^{(B)}, \sigma_1^{(A)} \cdot \sigma_{\phi}^{(B)}\right)|\Psi >.
\end{eqnarray}
Hence, we assign to each observable ${\cal O}$ on each path the corresponding eigenvalue of the operator ${\cal P}_o\left({\bf 1}, \sigma_1^{(A)}, \sigma_{\phi}^{(B)}, \sigma_1^{(A)} \cdot \sigma_{\phi}^{(B)}\right)$. That is,

\begin{equation}
\label{the_law2}
\left({\cal O}\right)_{cl}(\pm, \pm) = {\cal P}_o\left(+1, \left(\sigma_1^{(A)}\right)_{cl}(\pm, \pm), \left(\sigma_{\phi}^{(B)}\right)_{cl}(\pm, \pm), \left(\sigma_1^{(A)} \cdot \sigma_{\phi}^{(B)}\right)_{cl}(\pm, \pm)\right).
\end{equation}
Since the linear operators ${\cal P}_o\left({\bf 1}, \sigma_1^{(A)}, \sigma_{\phi}^{(B)}, \sigma_1^{(A)} \cdot \sigma_{\phi}^{(B)}\right)$ are not necessarily hermitic, their eigenvalues are, in general, complex numbers and, hence, also the c-values associated to these physical observables along the paths. 

The c-values assigned to physical observables according to these rules are actually their weak values for the corresponding post-selected states \cite{AAV,AV,AACV,aharonov}, 

\begin{equation}
\left({\cal O}\right)_{cl}(\pm, \pm) \equiv \frac{< \pm, \pm| {\cal P}_o | \Psi >}{< \pm, \pm| \Psi >} = \frac{< \pm, \pm| {\cal O} | \Psi >}{< \pm, \pm| \Psi >}.
\end{equation} 
Weak values of physical observables were first introduced as average values of post-selected weak measurements. Therefore, it is very tempting to try to interpret weak measurements as protocols to measure the values of physical observables along these paths.

In the particular case that we are considering here, in which both photons travel freely after being emitted, the hermitic operators describing physical observables in the Heisenberg picture do not evolve in time and, therefore, neither their c-values on paths do. More generally, operators describing physical observables evolve in time as ${\cal O}(t) \equiv e^{+ i \ {\cal H} \ t} \ {\cal O} \ e^{- i \ {\cal  H} \ t}$, where ${\cal H}$ is the hamiltonian of the system and time $t=0$ is set at the instant of post-selection, so that, their corresponding c-values on paths do also evolve, 

\begin{equation}
\left({\cal O}\right)_{cl}(t)(\pm, \pm) =  {\cal P}_{o(t)}\left(1, \left(\sigma_1^{(A)}\right)_{cl}(\pm, \pm), \left(\sigma_{\phi}^{(B)}\right)_{cl}(\pm, \pm), \left(\sigma_1^{(A)} \cdot \sigma_{\phi}^{(B)}\right)_{cl}(\pm, \pm)\right).
\end{equation}
From the equation 
\begin{equation}
\frac{d P_{o(t)}}{dt}|\Psi>=\frac{d {\cal O}(t)}{dt}|\Psi>= i [H,{\cal O}(t)]|\Psi>= i P_{[H,{\cal O}(t)]}|\Psi>
\end{equation}
we can obtain generic differential equations of motion for the pseudoclassical values on paths of any observable, 
\begin{equation}
\label{pseudo_equation}
\frac{d\left({\cal O}(t)\right)_{cl}}{dt}= i \left([H,{\cal O}(t)]\right)_{cl}.
\end{equation}

We have thus formally set a statistical interpretation of the quantum wavefunction (\ref{Bell_state}) as a mixture of non-interfering paths that obey pseudo-classical differential equations of motion. We showed at the end of the previous section that this statistical interpretation reproduces the quantum average values of all physical observables,  

\begin{equation}
\label{the_average}
< \Psi| {\cal O} |\Psi > =  < \Psi| P_{o}  |\Psi > = \sum_{\pm, \pm} \left|q_{\pm, \pm}\right|^2 \ \left({\cal O}\right)_{cl}(\pm, \pm),
\end{equation}
as well as all two-points quantum correlations,

\begin{equation}
\label{the_variance}
< \Psi| {\cal O}_1 \cdot {\cal O}_2 |\Psi > = < \Psi| P_{o_1}^{\dagger} \cdot P_{o_2} |\Psi > = \sum_{\pm, \pm} \left|q_{\pm, \pm}\right|^2 \left({\cal O}_1 \right)_{cl}^*(\pm, \pm) \ \left({\cal O}_2 \right)_{cl}(\pm, \pm).
\end{equation}

{\bf 3.} In order to demonstrate how the formalism of pseudo-classical paths works, we will explicitly show  in this section how to assign pseudo-classical values on paths to observables defined on the Hilbert space of two photons polarization states.

For example, in order to assign values to the observable $\sigma_1^{(B)}$ we rely on the linear relationship

\begin{eqnarray*}
\sigma_1^{(B)}|\Psi > \ =  \  -\sigma_1^{(A)}| \Psi > 
\end{eqnarray*} 
such that,
\begin{eqnarray*}
\left(\sigma_1^{(B)}\right)_{cl}(\pm, \pm) \ = \ -\left(\sigma_1^{(A)}\right)_{cl}(\pm, \pm). 
\end{eqnarray*}
The pseudo-classical values in the right hand side of this expression where defined in (\ref{assignments1},\ref{assignments1b}).\

Likewise, in order to assign values to $\sigma_2^{(B)}$ we notice that:
\begin{eqnarray*}
\sigma_2^{(B)}|\Psi > \ =  \  \sin^{-1}\left(\phi \right)\left[\sigma_{\phi}^{(B)} - \cos\left(\phi\right) \sigma_1^{(B)}\right]| \Psi > \ = \  \sin^{-1}\left(\phi \right)\left[\sigma_{\phi}^{(B)} + \cos\left(\phi\right) \sigma_1^{(A)}\right]| \Psi >,
\end{eqnarray*} 
which implies
\begin{eqnarray*}
\left(\sigma_2^{(B)}\right)_{cl}(\pm, \pm) \ = \    \sin^{-1}\left(\phi \right) \left[\left(\sigma_{\phi}^{(B)}\right)_{cl}(\pm, \pm) + \cos\left(\phi\right) \left(\sigma_1^{(A)}\right)_{cl}(\pm, \pm) \right]. 
\end{eqnarray*}
In order to assign values to the observable $\sigma_3^{(B)}$ we notice first that $\sigma_3^{(B)} = -i \sigma_1^{(B)} \sigma_2^{(B)}$
and, therefore,
\begin{eqnarray*}
\sigma_3^{(B)}|\Psi > = -i \sigma_1^{(B)} \sigma_2^{(B)}|\Psi > = i \sigma_2^{(B)} \sigma_1^{(B)}|\Psi > = -i \sigma_2^{(B)} 
\sigma_1^{(A)}|\Psi > = -i \sigma_1^{(A)} \sigma_2^{(B)}|\Psi > = \\
= -i \sin^{-1}\left(\phi \right)\left[\sigma_1^{(A)} \ \sigma_{\phi}^{(B)} + \cos\left(\phi\right) {\bf 1}\right]| \Psi >,
\end{eqnarray*}
which leads us to the following assignments:
\begin{eqnarray*}
\left(\sigma_3^{(B)}\right)_{cl}(\pm, \pm) & = & -i \sin^{-1}\left(\phi \right)\left[\left(\sigma_1^{(A)}\right)_{cl}(\pm, \pm) \cdot \left(\sigma_{\phi}^{(B)}\right)_{cl}(\pm, \pm) + \cos\left(\phi\right) \right]. 
\end{eqnarray*}
Any other observable defined on the Hilbert state of photon B can be written as a linear combination of $\sigma_1^{(B)}, \sigma_2^{(B)}, \sigma_3^{(B)},$ and the identity operator ${\bf 1}$ and, hence, its values on paths can be obtained from theirs.

\
In order to assign values to observables defined on the Hilbert space of photon A we proceed similarly. We have already given values to  $\sigma_1^{(A)}$. We can now exploit the perfect anti-correlation between the polarizations of photon A and photon B to assign values to $\sigma_2^{(A)}$ and $\sigma_3^{(A)}$. For example,

\begin{eqnarray*}
\sigma_2^{(A)}|\Psi > \ =  \  -\sigma_2^{(B)}| \Psi >,
\end{eqnarray*} 
which implies
\begin{eqnarray*}
\left(\sigma_2^{(A)}\right)_{cl}(\pm, \pm) \ = \ -\left(\sigma_2^{(B)}\right)_{cl}(\pm, \pm) = -\sin^{-1}\left(\phi \right) \left[\left(\sigma_{\phi}^{(B)}\right)_{cl}(\pm, \pm) + \cos\left(\phi\right) \left(\sigma_1^{(A)}\right)_{cl}(\pm, \pm) \right].
\end{eqnarray*}
Any other observable defined on the Hilbert state of photon A can be written as a linear combination of $\sigma_1^{(A)}, \sigma_2^{(A)}, \sigma_3^{(A)},$ and the identity operator ${\bf 1}$ and, hence, its values on paths can be obtained from theirs.

\

Finally, we can obtain the values on paths for any observable ${\cal O}$ defined on the Hilbert space of two photons polarization states by noticing that it can be written as ${\cal O} = \alpha_{00} \ {\bf 1} + \sum_{i=1,2,3} \ \alpha_{i0} \ \sigma_i^{(A)} +  \sum_{j=1,2,3} \ \alpha_{0j} \ \sigma_j^{(B)} + \sum_{i,j=1,2,3} \ \alpha_{i j} \ \sigma_i^{(A)} \cdot \sigma_j^{(B)}$ with $\alpha_{00}, \alpha_{i0}, \alpha_{0j}, \alpha_{i j} \in {\bf R}$ and then using the rules set above. For example,

\begin{equation}
\sigma_1^{(A)} \cdot \sigma_2^{(B)}|\Psi> = \sin^{-1}\left(\phi \right)\left[\sigma_1^{(A)} \ \sigma_{\phi}^{(B)} + \cos\left(\phi\right) {\bf 1}\right]| \Psi >.
\end{equation} 

\

In the limit $\phi \rightarrow 0$, which corresponds to choosing almost parallel directions to define the polarization of the two photons, only $|q_{+-}|^2$ and $|q_{-+}|^2$ are different from zero and, therefore, only paths $(+,-)$ and $(-,+)$ has non-zero (and equal) probabilities to happen. The pseudo-classical values of the photons polarizations in this limit can be easily obtained by noticing that:  

\begin{eqnarray*}
 \ \lim_{\phi \rightarrow 0} \frac{1 - \cos(\phi)}{\sin(\phi)} = 0.
\end{eqnarray*}
In particular, their polarizations along an arbitrary direction in the $XY$ plane $\sigma_\chi^{(A,B)} = \cos\left(\chi\right) \sigma_1^{(A,B)} + \sin\left(\chi\right) \sigma_2^{(A,B)}$ are given by: 

\begin{eqnarray*}
\left(\sigma_{\chi}^{(B)}\right)_{cl}(+, -)= -\left(\sigma_{\chi}^{(A)}\right)_{cl}(+, -) = - \cos\left(\chi\right), \\
\left(\sigma_{\chi}^{(B)}\right)_{cl}(-, +)= -\left(\sigma_{\chi}^{(A)}\right)_{cl}(-, +) = + \cos\left(\chi\right), \\
\end{eqnarray*}
which transform as classical vectors components. Besides the polarization along the $Z$ direction is zero on both paths,

\begin{eqnarray*}
\left(\sigma_{3}^{(B)}\right)_{cl}(+, -)= -\left(\sigma_{3}^{(A)}\right)_{cl}(+, -) = 0.
\end{eqnarray*}
 \

{\bf 4.} Actually, the freedom to build upon any complete set of commuting observables $\{\sigma_1^{(A)}, \sigma_{\phi}^{(B)}\}$ provides us with a continuous family of statistical representations of the same quantum state (\ref{Bell_state}) and we need to ask if all these different representations can describe a common unique physical {\it reality}. Indeed, the probabilities associated to each one of the four pseudo-classical paths (\ref{probabilities_123}) as well as the c-values of physical observables along each of them (\ref{pseudo_equation}) get modified under a transformation of the complete set of commuting observables used to describe the quantum state. In this section we explore the relationship between the statistical representations associated to different sets of commuting observables and in section 5 we show how they can be naturally understood as different coarse descriptions of a common unique finely resolved statistical representation. 

Let $\{{|\zeta_i>}\}_{i \in I}$ and $\{{|\xi_j>}\}_{j \in J}$ be families of common orthonormal eigenstates for two complete sets of commuting observables and

\begin{equation}
\left\{ \left({\cal O}_{\zeta}(t)\right)^{(i)}_{cl} = \frac{<\zeta_i|{\cal O}(t)|\Psi>}{<\zeta_i|\Psi>} \right\}_{i \in I}, \hspace{0.6in} 
\left\{ \left({\cal O}_{\xi}(t)\right)^{(j)}_{cl} = \frac{<\xi_j|{\cal O}(t)|\Psi>}{<\xi_j|\Psi>} \right\}_{j \in J}
\end{equation} 
be the pseudo-classical values of the observable ${\cal O}(t)$ along the paths associated to each of them. In these expressions is implicit the assumption that in both representations all paths have non zero probabilities to occur, $p_i=\left|<\zeta_i|\Psi> \right|^2 \neq 0$ for all $i \in I$ and $p_j = \left|<\xi_j|\Psi>\right|^2 \neq 0$ for all $j \in J$.

 It is straightfoward to obtain then the following relationship:

\begin{eqnarray*}
\left({\cal O}_{\zeta}(t)\right)^{(i \in I)}_{cl} = \frac{<\zeta_i|{\cal O}(t)|\Psi>}{<\zeta_i|\Psi>} = 
 \frac{\sum_{j \in J} <\zeta_i|\xi_j> <\xi_j|{\cal O}(t)|\Psi>}{<\zeta_i|\Psi>} = \\
= \sum_{j \in J} \frac{ <\zeta_i|\xi_j><\xi_j|\Psi>}{<\zeta_i|\Psi>} \frac{<\xi_j|{\cal O}(t)|\Psi>}{<\xi_j|\Psi>}= \\
= \sum_{j \in J} \frac{ <\zeta_i|\xi_j><\xi_j|\Psi>}{<\zeta_i|\Psi>} \left({\cal O}_{\xi}(t)\right)^{(j)}_{cl}.
\end{eqnarray*} 
Hence,  when we change the set of commuting observables used to represent the system the pseudo-classical values of a physical observable along the different paths transform linearly as the components of a vector in ${\bf C}^n$, where $n=4$ is the dimension of the Hilbert space of the system. The matrix of this transformation has complex coefficients:

\begin{equation}
\label{ext_probabilities}
{\widetilde p}_{j/i} =  \frac{ <\zeta_i|\xi_j><\xi_j|\Psi>}{<\zeta_i|\Psi>}  =  \frac{ <\Psi|\zeta_i><\zeta_i|\xi_j><\xi_j|\Psi>}{<\Psi|\zeta_i> <\zeta_i|\Psi>} \in {\bf C},
\end{equation}
such that:

\begin{eqnarray}
\label{the_formula}
\left({\cal O}_{\zeta}(t)\right)^{(i \in I)}_{cl} = \sum_{j \in J} {\widetilde p}_{j/i} \cdot \left({\cal O}_{\xi}(t)\right)^{(j)}_{cl}.
\end{eqnarray} 
This relationship can be rewritten as 
\begin{eqnarray}
\label{the_formula2}
0 = \sum_{j \in J} {\widetilde p}_{j/i} \cdot \left(\left({\cal O}_{\xi}(t)\right)^{(j)}_{cl} - \left({\cal O}_{\zeta}(t)\right)^{(i \in I)}_{cl}\right),
\end{eqnarray} 
by noticing that 

\begin{equation}
\label{Bayes}
 \sum_{j \in J} {\widetilde p}_{j/i} = \sum_{j \in J}  \frac{ <\Psi|\zeta_i><\zeta_i|\xi_j><\xi_j|\Psi>}{<\Psi|\zeta_i> <\zeta_i|\Psi>} =  \frac{ <\Psi|\zeta_i><\zeta_i|\Psi>}{<\Psi|\zeta_i> <\zeta_i|\Psi>} = 1.
\end{equation}
That is, for all physical observables ${\cal O}(t)$ the vector ${\vec \Omega}_{o(t)}^{(i \in I)} \equiv \left(\left({\cal O}_{\xi}(t)\right)^{(j)}_{cl}  - \left({\cal O}_{\zeta}(t)\right)^{(i \in I)}_{cl} \right)_{j \in J}  \in {\bf C}^n$ is contained within the hyperplane of dimension $n-1$ orthogonal to the vector $\left({\widetilde p}^*_{j/i} \right)_{j \in J} \in {\bf C}^n$. Moreover, the vectors $\left\{{\vec \Omega}_{\sigma_1^{(A)}}^{(i \in I)}, {\vec \Omega}_{\sigma_{\phi}^{(B)}}^{(i \in I)}, {\vec \Omega}_{\sigma_1^{(A)}\cdot\sigma_{\phi}^{(B)}}^{(i \in I)} \right\}$ associated to the linear generators of a commutative algebra (\ref{the_law}) of all the operators with common eigenvectors (excluiding from them the vector ${\vec \Omega}_{\bf 1}^{(i \in I)}=0$ associated to the identity operator) form a basis in this hyperplane \footnote{The proof is straightforward: the linear relationship $\lambda_1 {\vec \Omega}_{\sigma_1^{(A)}}^{(i \in I)} + \lambda_2 {\vec \Omega}_{\sigma_{\phi}^{(B)}}^{(i \in I)} + \lambda_3 {\vec \Omega}_{\sigma_1^{(A)}\cdot\sigma_{\phi}^{(B)}}^{(i \in I)} = 0$ means that $\lambda_1 \ \cdot <\xi_j|\sigma_1^{(A)}|\Psi> + \ \lambda_2 \ \cdot <\xi_j|\sigma_{\phi}^{(B)}|\Psi> + \ \lambda_3 \  \cdot <\xi_j|\sigma_1^{(A)} \cdot \sigma_{\phi}^{(B)}|\Psi> =  \omega \ \cdot <\xi_j|\Psi>$ for all vectors $\{|\xi_j>\}_{j \in J}$, with $\omega = 
\lambda_1 \frac{<\zeta_i|\sigma_1^{(A)}|\Psi>}{<\zeta_i|\Psi>} + \lambda_2 \frac{<\zeta_i|\sigma_{\phi}^{(B)}|\Psi>}{<\zeta_i|\Psi>} + \lambda_3  \frac{<\zeta_i|\sigma_1^{(A)} \cdot \sigma_{\phi}^{(B)}|\Psi>}{<\zeta_i|\Psi>} \in {\bf C}$. That is,  $\lambda_1 \ \cdot \sigma_1^{(A)}|\Psi> + \ \lambda_2 \ \cdot \sigma_{\phi}^{(B)}|\Psi> + \ \lambda_3 \  \cdot \sigma_1^{(A)} \cdot \sigma_{\phi}^{(B)}|\Psi> =  \omega \ \cdot |\Psi>$, which implies $\lambda_1=\lambda_2=\lambda_3=0$ because the vectors  $\sigma_1^{(A)}|\Psi>$, $\sigma_{\phi}^{(B)}|\Psi>$, $\sigma_1^{(A)} \cdot \sigma_{\phi}^{(B)}|\Psi>$ and $|\Psi>$ are linearly independent, see (\ref{the_law}).}.

The complex coefficients  ${\widetilde p}_{j/i}$ in (\ref{the_formula}) can also be understood as conditional pseudo-probabilities in terms of an extended version of Bayes law, as in addition to (\ref{Bayes}) they also fullfill the constraint 

\begin{equation}
 \sum_{i \in I} p_{i} \cdot {\widetilde p}_{j/i} = \sum_{i \in I} <\Psi|\zeta_i> <\zeta_i|\Psi> \cdot \frac{ <\Psi|\zeta_i><\zeta_i|\xi_j><\xi_j|\Psi>}{<\Psi|\zeta_i> <\zeta_i|\Psi>} = <\Psi|\xi_j><\xi_j|\Psi> = p_{j \in J}.
\end{equation}
The extension of the standard probability theory to the case in which pseudo-probabilities can take values outside the real interval $\left[0, 1\right]$ has been largely discussed in the literature since the very early stages of the development of Quantum Mechanics \cite{Wigner}. In particular, the complex pseudo-probabilities ${\widetilde p}_{j/i}$ defined in (\ref{ext_probabilities}) were first  introduced in a different context in \cite{Kirkwood}. A formal axiomatic formulation of extended pseudo-probability models was first laid down in \cite{Cox, Bartlett} and more recently in \cite{Burgin}. In general, it is widely recognized that pseudo-probabilities outside the interval $\left[0, 1\right]$ can be formally associated to events that are not physical, as long as probabilities of all physical events lay within this interval. This is indeed the case of the conditional complex pseudo-probabilities ${\widetilde p}_{j/i}$, whose {\it naive} interpretation would be the probability of obtaining the set $j \in J$ of eigenvalues of a given complete family of strongly measured commuting observables given that the set $i \in I$ of eigenvalues of a different complete family of commuting observables, not commuting with the former, had been previously strongly measured (without altering the state of the system). 
\\

The argument that follows tries to offer a more intuitive understanding of the physical meaning of these complex conditional pseudo-probabilities. It shows that pseudo-probabilities are required when trying to statistically describe coarsely tested physical systems and, therefore, it implies that the pseudo-classical paths that we have built above are indeed coarse descriptions of an underlying finely resolved hidden reality. In order to make our argument clearer we first present it with the help of a simple example in the rest of this section. Then, in section 5, we explicitly show how the pseudo-classical paths built in sections 2 and 3 to describe Bohm's two photons system in their singlet polarization state can be understood as coarse statistical descriptions of finely resolved hidden paths, grouped into large subsets by imposing post-selection conditions on certain sets of physical observables.  

We consider a random real variable $z_{i,j} \in {\bf R}$ defined on a probability space consisting of $n \times n$ possible single events $\{(i,j)\}_{i,j=1,...,n}$, each one occurring with probability $p_{i,j}$. We assume that we can prepare infinitely many realizations of this probabilistic model. On the other hand, we also assume that only two kinds of experimental tests can be performed on this system: strong tests return either the row $i=1,...,n$ or the column $j=1,..,n$ of each particular event, while weak tests return low precision estimations of the variable $z_{i,j}$ at that event.       
\\

\hspace{1.4in}
\begin{tabular}{|r|r|r|r|r|r|}
\hline
$z_{1,1}$&$z_{1,2}$&$.........$&$.........$&$.........$&$z_{1,n}$\\
\hline
$z_{2,1}$&$z_{2,2}$&$.........$&$.........$&$.........$&$z_{2,n}$\\
\hline
$.........$&$.........$&$.........$&$.........$&$.........$&$........$\\
\hline
$.........$&$.........$&$.........$&$z_{i,j}$&$.........$&$........$\\
\hline
$.........$&$.........$&$.........$&$.........$&$.........$&$........$\\
\hline
$z_{n,1}$&$z_{n,2}$&$.........$&$.........$&$.........$&$z_{n,n}$\\
\hline
\end{tabular}

\[ \mbox{Figure 1. Schematic representation of the finely described probabilistic toy model.} \] \\
Under such conditions, we can measure with any desired precision the statistics of rows, $p^{row}_i=\sum_{j=1,..,n} p_{i,j}$, and columns,  $p^{col}_j=\sum_{i=1,..,n} p_{i,j}$. We can also measure with any desired precision the average value of the random variable along any row $Z^{row}_{(i=1,...,n)} = \frac{1}{p_i} \sum_{j=1,...,n} p_{i,j} \cdot z_{i,j}$ or column $Z^{col}_{(j=1,...,n)} = \frac{1}{p_j} \sum_{i=1,...,n} p_{i,j} \cdot z_{i,j}$, by performing a weak measurement followed by a strong measurement on infinitely many realizations of the probabilistic model and post-selecting only those events that happen to fall on the required row or column. Notwithstanding, we cannot exploit the same strategy to get precision measurements of the probability $p_{i,j}$ or the value of the random variable $z_{i,j}$ at any single event.
\\   
\\


\begin{tabular}{|r|}
\hline
$\hspace{1.1in}  Z^{row}_1    \hspace{1.1in}$\\
\hline
$\hspace{1.1in}   Z^{row}_2    \hspace{1.1in}$\\
\hline
$\hspace{1.1in}   ....    \hspace{1.1in}$\\
\hline
$\hspace{1.1in}  Z^{row}_ i    \hspace{1.1in}$\\
\hline
$\hspace{1.1in}   ....    \hspace{1.1in}$\\
\hline
$\hspace{1.1in}  Z^{row}_n    \hspace{1.1in}$\\
\hline
\end{tabular}
\hspace{0.5in}
\begin{tabular}{|r|r|r|r|r|r|}
\hline
 &  & & & & \\
 &  & & & & \\
 &  & & & & \\
$\hspace{0.12in} Z^{col}_1 \hspace{0.12in}$    &$\hspace{0.12in} Z^{col}_2 \hspace{0.12in}$   &$\hspace{0.07in} ...... \hspace{0.07in}$   &$\hspace{0.12in} Z^{col}_j \hspace{0.12in}$   &$\hspace{0.07in} ...... \hspace{0.07in}$   &$\hspace{0.12in} Z^{col}_n \hspace{0.12in}$\\
 &  & & & & \\
 &  & & & & \\
\hline
\end{tabular}
 
\[ \mbox{Figure 2. Experimentally accessible coarse subsets within the probabilistic toy model.} \] \\
\\ \\
If we now try to express  the average values of the random variable along rows as a linear mean of its average values along columns, 

\begin{equation}
\label{STAR}
Z^{row}_i = \sum_{j=1,...,n} {\widetilde p}_{j/i} \cdot Z^{col}_j,
\end{equation}
we cannot any longer require the real linear coefficients ${\widetilde p}_{j/i}$ to lay within the unit interval $[0, 1]$ if they are required to fulfill also

\begin{equation}
\label{manifold}
\sum_{j} {\widetilde p}_{j/i} = 1,  \hspace{1.2in} \sum_{i} p^{row}_i \cdot {\widetilde p}_{j/i} = p^{col}_j,
\end{equation}
because $Z^{col}_j$ is not necessarily equal to $z_{i,j}$. Nonetheless, if we do not demand from the linear coefficients ${\widetilde p}_{j/i}$ to belong to the unit interval, the linear equation (\ref{STAR}) always admits a solution $\left({\widetilde p}_{j/i}\right)_{i,j=1,...,n} \in {\cal M}({\bf R}, n)$ that fulfills the constraints (\ref{manifold}), except for the singular cases in which $(Z^{col}_j)_{j=1,...,n} \propto {\bf 1}$ and $(Z^{col}_i)_{i=1,...,n} \neq (Z^{col}_j)_{j=1,...,n}$. Thus,  we conclude that conditional pseudo-probabilities outside the unit interval are required in order to describe statistical systems that can only be coarsely tested. \\

{\bf 5.} The arguments presented in section 4 suggest that the sets of pseudo-classical paths that we defined in the previous sections 1,2 and 3 associated  to each complete set of commuting observables may indeed be different coarse descriptions of a common underlying set of finely resolved paths. In this section we turn on to the task of building this underlying set of finely resolved paths, which constitutes an explicit pseudo-classical statistical model of {\it hidden} variables that reproduces the expectation values, correlations and weak values of all quantum observables in Bohm's system of two photons in their singlet polarization state. 

We consider a statistical system whose phase space consists of an infinitely large number of equally probable states distributed over the unit circle ${\cal S}_1$. An observer that strongly measures the polarization of either one of the photons, say photon $A$, fixes a reference direction $\Omega_0$ on this circle. In this particular frame the number density distribution of states over the circle is given by

\begin{equation}
\label{number_density_distribution_of_states}
g\left(\omega\right) = \frac{1}{4} \left|\sin\left(\omega\right)\right|,
\end{equation}  
where $\omega \in \left[-\pi, \pi\right)$ is an angular coordinate defined with respect to the chosen reference direction. 
This statement about the dependence of the number density distribution of polarization states on the direction of observation is crucial in our proposal to solve the EPR paradox.

The observer of photon $B$ may measure its polarization along a different direction $\Omega'_0 = \Omega_0 + \Delta \Omega_0$, shifted by an angle $\Delta \Omega_0  \in [0, \pi)$ with respect to the reference direction set by the observer of photon $A$ \footnote{We can assume that $\Delta \Omega_0  \in [0, \pi)$ without any loss of generality because in the case that $\Delta \Omega_0  \in [-\pi, 0)$ we could change the roles of photons $A$ and $B$.}. This new observer parameterizes the {\it hidden} phase space $S_1$ in terms of his own angular coordinate $\omega' \in \left[-\pi, \pi\right)$,  
defined with respect to the reference direction that he chose. The transformation law  $\omega \rightarrow \omega'$ that relates the coordinates of a given {\it hidden} state as defined by the observers of photons $A$ and $B$ is fixed by the symmetry demand that they both see the same number density distribution of states (\ref{number_density_distribution_of_states}) and the following boundary conditions  

\begin{eqnarray*}
\begin{array}{ccccccc}
\omega & = & 0 & \longrightarrow &  \omega' & = & \Delta \Omega_0, \\
\omega & = & \Delta \Omega_0 & \longrightarrow &  \omega' & = & 0, \\
\omega & = & \pm \pi & \longrightarrow &  \omega' & = & \Delta \Omega_0-\pi, \\
\omega & = & \Delta \Omega_0-\pi & \longrightarrow &  \omega' & = & \pm\pi.
\end{array}
\end{eqnarray*}
Therefore,

\begin{eqnarray}
\label{Oaknin_transformation}
\omega \rightarrow \omega' =  
\left\{
\begin{array}{cccccccc}
S(\omega) \cdot \mbox{acos}\left(-\cos(\Delta\Omega_0) - \cos(\omega) - 1 \right), \hspace{0.15in} & \mbox{if}  & \hspace{0.15in} -\pi & \le & \omega & < & \Delta\Omega_0-\pi, \\
S(\omega) \cdot \mbox{acos}\left(+\cos(\Delta\Omega_0) + \cos(\omega) - 1 \right), \hspace{0.15in} & \mbox{if}  & \hspace{0.15in} \Delta\Omega_0-\pi & \le & \omega & < & 0, \\
S(\omega) \cdot \mbox{acos}\left(+\cos(\Delta\Omega_0) - \cos(\omega) + 1 \right), \hspace{0.15in} & \mbox{if}  & \hspace{0.15in} 0 & \le & \omega & < & \Delta \Omega_0, \\
S(\omega) \cdot \mbox{acos}\left(-\cos(\Delta\Omega_0) + \cos(\omega) + 1 \right), \hspace{0.15in} & \mbox{if}  & \hspace{0.15in} \Delta \Omega_0 & \le & \omega & < & +\pi, \\
\end{array}
\right.
\end{eqnarray}
where 
\begin{eqnarray*}
S(\omega) =-\mbox{sign}((\omega - \Delta \Omega_0) \mbox{mod} ([-\pi, \pi))),
\end{eqnarray*}
and the function $y=\mbox{acos}(x)$ is defined in his main branch, such that $y \in [0, \pi]$ while $x \in [-1, +1]$. It is strightforward to prove that for any value of $\Delta \Omega_0 \in [0, \pi]$  this transformation leaves the number density distribution of states (\ref{number_density_distribution_of_states}) invariant, as

\begin{equation}
d\omega' \ |\sin(\omega')| = |d(\cos(\omega'))| = |d(\cos(\omega))| = d\omega \ |\sin(\omega)|.
\end{equation}
Indeed, the transformation law (\ref{Oaknin_transformation}) defines a group of symmetry operations in the {\it hidden} phase space $S_1$ that leaves the number density distribution  (\ref{number_density_distribution_of_states}) invariant \footnote{This transformation law for the angular coordinate of the {\it hidden} states under rotations of the angular reference direction reminds the Lorentz transformation that relates two inertial frames moving with relativistic relative velocity $v$. Indeed, we can draw a further similarity between the symmetry invariace of the number density distribution of states (\ref{number_density_distribution_of_states}) under rotations and the invariance of the number density distribution of photons which are emitted isotropically in the proper frame of a decaying scalar particle as described by observers moving with relativistic velocity with respect to it in different angular directions.}. In particular, when $\Delta\Omega_0=0$ the transformation is simply a parity operation, $\omega \rightarrow \omega'=-\omega$, while for $\Delta\Omega_0=\pi$ the angular coordinate transform as $\omega \rightarrow \omega'=\pi-\omega$ if $\omega \ge 0$ and $\omega \rightarrow \omega'=-\pi-\omega$ if $\omega < 0$.

We now define a strong measurement of the polarization of photon A along the reference direction $\Omega_0$ as a test of the sign of the angular coordinate $\omega$ of the {\it hidden} state of the system in the fixed frame. That is,  

\begin{equation}
\label{partition_1A}
S_{\Omega_0}^{(A)}(\omega) =  \left\{
\begin{array}{ccccc}
+1,& \hspace{0.2in} if \hspace{0.2in} \omega \in & [& 0,& \pi), \\
-1,&  \hspace{0.2in} if \hspace{0.2in} \omega \in & [&-\pi,& 0),
\end{array}
\right.
\end{equation}  
the output of the strong measurement is positive if the system happen to be in a {\it hidden} state with positive angular coordinate and it is negative otherwise. Obviously, each of one of the two possible outputs happen with probability $1/2$. 

Similarly, a strong measurement of  the polarization of photon B along the reference direction $\Omega'_0$ tests the sign of the angulat coordinate $\omega'$ of the {\it hidden} state. That is,

\begin{equation}
\label{partition_2B}
S_{\Omega'_0}^{(B)}(\omega') =  \left\{
\begin{array}{cccccc}
+1,& \hspace{0.2in} if \hspace{0.2in} \omega' \in & [& 0,& \pi), \\
-1,&  \hspace{0.2in} if \hspace{0.2in} \omega' \in & [&-\pi,& 0).
\end{array}
\right.
\end{equation} 
By using the transformation law (\ref{Oaknin_transformation}) it is straightforward to check that, 

\begin{equation}
S_{\Omega'_0}^{(B)}(\omega) =  \left\{
\begin{array}{cccccc}
+1,& \hspace{0.2in} if \hspace{0.2in} \omega \in & [\Delta \Omega_0-\pi,\Delta \Omega_0), \\
-1,&  \hspace{0.2in} if \hspace{0.2in} \omega \in & [-\pi, \Delta \Omega_0-\pi) \bigcup [\Delta \Omega_0, \pi).
\end{array}
\right.
\end{equation} 
%
Therefore, the four possible outputs of the two strong measurements define a partition of the {\it hidden} phase space into four coarse subsets,

\begin{eqnarray}
\label{four_subsets}
(S_{\Omega_0}^{(A)}=+1; \ S_{\Omega'_0}^{(B)}=+1) & \Longleftrightarrow & \omega \in [0, \Delta \Omega_0) \\
(S_{\Omega_0}^{(A)}=+1; \ S_{\Omega'_0}^{(B)}=-1) & \Longleftrightarrow & \omega \in [\Delta \Omega_0, \pi) \\
(S_{\Omega_0}^{(A)}=-1; \ S_{\Omega'_0}^{(B)}=+1) & \Longleftrightarrow & \omega \in [\Delta \Omega_0-\pi, 0) \\
(S_{\Omega_0}^{(A)}=-1; \ S_{\Omega'_0}^{(B)}=-1) & \Longleftrightarrow & \omega \in [-\pi, \Delta \Omega_0-\pi),
\end{eqnarray}
whose probabilities to happen are: 

\begin{eqnarray*}
p\left(S_{\Omega_0}^{(A)}=+1, S^{(B)}_{\Omega'_0}=+1\right) & =  & \int_0^{\Delta \Omega_0} g(\omega) \ d\omega \hspace{0.15in} = \ \frac{1}{4}\left(1 - \cos (\Delta \Omega_0)\right), \\
p\left(S_{\Omega_0}^{(A)}=+1, S^{(B)}_{\Omega'_0}=-1\right)  & =  & \int_{\Delta \Omega_0}^{\pi} g(\omega) \ d\omega \hspace{0.21in} = \ \frac{1}{4}\left(1 + \cos (\Delta \Omega_0)\right), \\
p\left(S_{\Omega_0}^{(A)}=-1, S^{(B)}_{\Omega'_0}=+1\right) & = & \int_{\Delta \Omega_0-\pi}^{0} g(\omega) \ d\omega \hspace{0.06in} = \ \frac{1}{4}\left(1 + \cos (\Delta \Omega_0)\right), \\
p\left(S_{\Omega_0}^{(A)}=-1, S^{(B)}_{\Omega'_0}=-1\right) & = & \int_{-\pi}^{\Delta \Omega_0-\pi} g(\omega) \ d\omega = \ \frac{1}{4}\left(1 - \cos (\Delta \Omega_0)\right),
\end{eqnarray*}
%
which reproduce the crossed probabilities (\ref{probabilities_123}) predicted by Quantum Mechanics for Bohm's two photons system in their singlet polarization state.  

In order to understand how this model avoids the constraints set up by Bell's theorem we try now to rewrite for it the general proof that we provided in section 1. We get the following inequality: 

\begin{eqnarray*}
\label{Bell_inequality_mebutal}
\left| E(S_{\Omega_0}, S_{\Omega'_0}) - E(S_{\Omega_0}, S_{\Omega''_0}) \right| & = & 4 \int_{\Omega''_0-\Omega_0}^{\Omega'_0-\Omega_0} \ d\omega \ g(\omega) \le   1 + 2 \int_{\Omega''_0-\Omega_0}^{\Omega'_0-\Omega_0} \ d\omega \ g(\omega),
\end{eqnarray*}
where we have assumed without any loss of generality that $\Omega'_0-\Omega_0, \Omega''_0-\Omega_0 \ge 0$ and $\Omega'_0-\Omega_0 \ge \Omega''_0-\Omega_0$. But the last step in the proof that appears in section 1 cannot be done for our model, because $2 \int_{\Omega''_0-\Omega_0}^{\Omega'_0-\Omega_0} \ d\omega \ g(\omega) = 2 \int_0^{\mbox{acos}(\cos({\Omega''_0-\Omega_0})-\cos({\Omega'_0-\Omega_0}))} \ d\omega' \ g(\omega')$  is not necessarily equal to

\begin{equation}
E(S_{\Omega''_0}, S_{\Omega'_0}) = 2 \int_0^{\Omega'_0-\Omega''_0} \ d\omega' \ g(\omega').
\end{equation}
\

This situation is somewhat analogous to what we get when we try to describe the isotropic decay of a massive scalar particle into two identical photons a the eyes of a family of inertial observers moving with respect to it with relativistic velocity along different angular directions. In order to make the analogy clearer we consider only those pairs of photons that are emitted within a given plane and observers that move also within the same plane. Then, the number of photons that an observer moving along direction $\Omega_0$ sees moving between angles $\Omega''_0-\Omega_0$ and  $\Omega'_0-\Omega_0$ defined with respect to it is not necessarily equal to the number of photons that an observer moving along direction $\Omega''_0$
(as defined by the former observer) would see moving within an angle $\Omega'_0-\Omega''_0$ as this new observer defines it.

In this {\it hidden} phase space of finely resolved states we now define the polarization properties of each photon with respect to the reference direction set by the observer looking at it, as follows:  

\begin{eqnarray*}
s_1(\omega) & = & \left\{ 
\begin{array}{ccccc}
\ \ +1, \ \ & \hspace{0.2in} \omega \in & [& 0,& \pi), \\
\ \ -1, \ \ &  \hspace{0.2in} \omega \in & [&-\pi,& 0).
\end{array}
\right. \\
s_2(\omega) & = & \left\{ 
\begin{array}{ccccc}
-\frac{1}{\tan \omega},& \hspace{0.2in} \omega \in & [& 0,& \pi), \\
+\frac{1}{\tan \omega},&  \hspace{0.2in} \omega \in & [&-\pi,& 0).
\end{array}
\right. \\
s_3(\omega) & = & \left\{ 
\begin{array}{ccccc}
+\frac{i}{\tan \omega},& \hspace{0.2in} \omega \in & [& 0,& \pi), \\
+\frac{i}{\tan \omega},&  \hspace{0.2in} \omega \in & [&-\pi,& 0).
\end{array}
\right.,
\end{eqnarray*}  
where $s_1$ denotes the polarization component along the reference direction set by the observer, $s_3$ denotes the polarization component along the photon's travelling direction and $s_2$ denotes the polarization component along their ortoghonal, right-oriented direction. The polarization of the photon along any other direction is defined linearly with respect to these three. For example:

\begin{eqnarray*}
s_{\Delta \Omega_0}(\omega) = \cos(\Delta \Omega_0) \cdot s_1(\omega) + 
\sin(\Delta \Omega_0) \cdot s_2(\omega) = \hspace{2.0in} \\ \hspace{2.0in}
= \left\{ 
\begin{array}{ccccc}
+\cos(\Delta \Omega_0)  - \frac{\sin(\Delta \Omega_0)}{\tan \omega}=+\frac{\sin(\omega - \Delta \Omega_0)}{\sin \omega},& \hspace{0.2in} \omega \in & [& 0,& \pi), \\
- \cos(\Delta \Omega_0)  +\frac{\sin(\Delta \Omega_0)}{\tan \omega}=-\frac{\sin(\omega - \Delta \Omega_0)}{\sin \omega},&  \hspace{0.2in} \omega \in & [&-\pi,& 0),
\end{array}
\right.
\end{eqnarray*} 
such that $s_{\Delta \Omega_0}(\Delta \Omega_0) = 0$. 
This situation in which the values of physical observables are defined within the frame of reference in which we choose to describe the system can also be found in Special Relativity. For example, two observables moving with relative velocity would not assign the same value to the component of an electric field along the direction of their relative motion.   

We now define weak measurements as low precision tests of the polarization components of the photons in this finely resolved phase space, followed by strong measurements of any complete set of commuting observables  $\{\sigma^{(A)}_{\Omega_0}, \sigma^{(B)}_{\Omega'_0}\}$.  By repeating the weak measurement on many identically prepared systems we can obtain, with high precision, the average value of any polarization component on each of the four different coarse subsets of the phase space (\ref{four_subsets}) defined by the output $S^{(A)}_{\Omega_0}=\pm1, \ S^{(B)}_{\Omega'_0}=\pm 1$ of the two strong measurements. It is straightforward to test that these average values reproduce the weak values of quantum observables as defined in Quantum Mechanics: \\

\begin{eqnarray*} 
\frac{\int_0^{\Delta \Omega_0} d\omega \ g(\omega) \ s_I^{(A)}(\omega)}{\int_0^{\Delta \Omega_0} d\omega \ g(\omega)}
 & = & \
\frac{<S_{\Omega_0}^{(A)}=+1; S_{\Omega'_0}^{(B)}=+1|\sigma_I^{(A)}|\Psi>}{<S_{\Omega_0}^{(A)}=+1; S_{\Omega'_0}^{(B)}=+1|\Psi>}, \\
\frac{\int_{\Delta \Omega_0}^{\pi} d\omega \ g(\omega) \ s_I^{(A)}(\omega)}{\int_{\Delta \Omega_0}^{\pi} d\omega \ g(\omega)}
 & = & \
\frac{<S_{\Omega_0}^{(A)}=+1; S_{\Omega'_0}^{(B)}=-1|\sigma_I^{(A)}|\Psi>}{<S_{\Omega_0}^{(A)}=+1; S_{\Omega'_0}^{(B)}=-1|\Psi>}, \\
\frac{\int_{\Delta \Omega_0-\pi}^0 d\omega \ g(\omega) \ s_I^{(A)}(\omega)}{\int_{\Delta \Omega_0-\pi}^0 d\omega \ g(\omega)}
 & = &
\frac{<S_{\Omega_0}^{(A)}=-1; S_{\Omega'_0}^{(B)}=+1|\sigma_I^{(A)}|\Psi>}{<S_{\Omega_0}^{(A)}=-1; S_{\Omega'_0}^{(B)}=+1|\Psi>}, \\
\frac{\int_{-\pi}^{\Delta \Omega_0-\pi} d\omega \ g(\omega) \ s_I^{(A)}(\omega)}{\int_{-\pi}^{\Delta \Omega_0-\pi} d\omega \ g(\omega)}
 & = &
\frac{<S_{\Omega_0}^{(A)}=-1; S_{\Omega'_0}^{(B)}=-1|\sigma_I^{(A)}|\Psi>}{<S_{\Omega_0}^{(A)}=-1; S_{\Omega'_0}^{(B)}=-1|\Psi>}.
\end{eqnarray*}
Hence, the four coarse paths resulting from the partition of the {\it hidden} phase space into subspaces according to the outputs of the strong measurement of the complete set of commuting observables $\{\sigma^{(A)}_{\Omega_0}, \sigma^{(B)}_{\Omega'_0}\}$ are identical to the paths that we introduced in the previous sections 1,2,3 associated to these observables. Furthermore, our model suggests that we can gain access to the actual values of the polarization components at each single {\it hidden} states by deriving the weak values of physical observables with respect to the angular shift $\Delta \Omega_0$. \\


In the general case in which either of the photons interacts with the external world we define the dynamics of each state in the {\it hidden} phase space
as we did above in sections 1,2, through the Heisenberg description of the time evolution of the quantum operators. This definition guarantees the locality of the model: the equality ${\cal O}_A(t) = e^{+ i H_B t} {\cal O}_A e^{-i H_B t} = {\cal O}_A$ implies that once a frame of reference has been chosen the actual values assigned to physical observables of one of the photons do not change as a result of an interaction of the other photon with the external world. Thus, the {\it hidden} variables model that we have built is an explicit pseudo-classical statistical description of Bohm's two photons system.
\\

\
\\
{\bf 6.} Let us now show how the statistical description of the quantum wavefunction that we have introduced can solve the EPR paradox. We consider a measuring device that at time $t=0$ interacts very briefly with one of the photons of the singlet state (\ref{Bell_state}) and performs a strong measurement of its polarization along the X-direction.  For the sake of simplicity we assume that the measuring device is an additional photon whose polarization is initially set in the state $| \downarrow >^{(*)}$.
Hence, the composite system is described by the wave function:

\begin{equation}
\label{the_enlarged_state}
|{\widetilde \Psi}> = | \downarrow >^{(*)} \otimes |\Psi> =  | \downarrow >^{(*)} \otimes \frac{1}{\sqrt{2}} \left( | \uparrow >^{(A)} \otimes  | \downarrow >^{(B)} -  | \downarrow >^{(A)} \otimes  | \uparrow >^{(B)} \right),
\end{equation}
and its dynamics by the time dependent unitary operator:

\begin{equation}
\label{the_dynamics}
U(t) = e^{- i H t} = 
\left\{
\begin{array}{cc}
{\bf 1}, & t < 0\\
-\sigma_1^{(*)} \otimes \frac{1}{2} \left({\bf 1} + \sigma_1 \right)^{(A)} \otimes {\bf 1}^{(B)} - \sigma_3^{(*)} \otimes \frac{1}{2} \left({\bf 1} - \sigma_1 \right)^{(A)} \otimes {\bf 1}^{(B)} , & t \ge 0.
\end{array}
\right. \\
\end{equation} \\
This transformation correlates the polarization of photon A along the X-direction with the polarization of the measuring device along the Z-direction: 

\begin{eqnarray*}
U(t) |{\widetilde \Psi}>= 
\left\{
\begin{array}{cc}
 |{\widetilde \Psi}>, & t < t_0\\
\frac{1}{\sqrt{2}} |\uparrow>^{(*)} \otimes |x+>^{(A)} \otimes |x->^{(B)} + \frac{1}{\sqrt{2}} |\downarrow>^{(*)} \otimes |x->^{(A)} \otimes |x+>^{(B)}, & t \ge t_0
\end{array}
\right.
\end{eqnarray*}
where $|x\pm> \equiv \frac{1}{\sqrt{2}} \left(|\uparrow> \pm |\downarrow>\right)$ are the eigenstates of the polarization operator $\sigma_1$.

Operators defined on the Hilbert space of photon B commute with the unitary transformation $U(t)$ and, therefore, they do not get modified by the interaction of photon A with the measuring device (as it should be expected from locality).

\begin{equation}
\label{3stars}
\sigma_{1,2,3}^{(B)}(t) = U(t)^{\dagger} \cdot \sigma_{1,2,3}^{(B)} \cdot U(t) = \sigma_{1,2,3}^{(B)}, \hspace{0.8in} t \in {\bf R}.
\end{equation}

The operator that describe the polarization of photon A along the X-direction also commutes with the unitary transformation $U(t)$ and, therefore, neither it does get modified by its interaction with the measuring device:

\begin{equation}
\sigma_{1}^{(A)}(t) = U(t)^{\dagger} \cdot \sigma_{1}^{(A)} \cdot U(t) = \sigma_{1}^{(A)}, \hspace{0.8in} t \in {\bf R}.
\end{equation} 
Nevertheless, the interaction with the measuring device does modify the polarization of photon A along any other direction: 

\begin{eqnarray}
\sigma_{2}^{(A)}(t) = U(t)^{\dagger} \cdot \sigma_{2}^{(A)} \cdot U(t) =
\left\{
\begin{array}{cc}
\sigma_2^{(A)}, & t < 0 \\
 \sigma_{2}^{(*)} \otimes \sigma_{3}^{(A)} \otimes {\bf 1}^{(B)}, \hspace{0.6in} & t \ge 0.
\end{array}
\right. 
\end{eqnarray} 

\begin{eqnarray}
\sigma_{3}^{(A)}(t) = U(t)^{\dagger} \cdot \sigma_{3}^{(A)} \cdot U(t) =
\left\{
\begin{array}{cc}
\sigma_3^{(A)}, & t < 0 \\
-\sigma_{2}^{(*)} \otimes \sigma_{2}^{(A)} \otimes {\bf 1}^{(B)}, \hspace{0.6in} & t \ge 0.
\end{array}
\right. 
\end{eqnarray} 

Similarly, the polarization components of the measuring device gets also transformed by its interaction with photon A:

\begin{eqnarray}
\sigma_{1}^{(*)}(t) = U(t)^{\dagger} \cdot \sigma_{1}^{(*)} \cdot U(t) =
\left\{
\begin{array}{cc}
\sigma_1^{(*)}, & t < 0 \\
 \sigma_{1}^{(*)} \otimes \sigma_{1}^{(A)} \otimes {\bf 1}^{(B)}, \hspace{0.6in} & t \ge 0.
\end{array}
\right. 
\end{eqnarray} 

\begin{equation}
\sigma_{2}^{(*)}(t) = U(t)^{\dagger} \cdot \sigma_{2}^{(*)} \cdot U(t) =
\left\{
\begin{array}{cc}
\sigma_2^{(*)}, & t < 0 \\
 -\sigma_{2}^{(*)},  & t \ge 0.
\end{array}
\right. 
\end{equation} 

\begin{eqnarray}
\sigma_{3}^{(*)}(t) = U(t)^{\dagger} \cdot \sigma_{3}^{(*)} \cdot U(t) =
\left\{
\begin{array}{cc}
\sigma_3^{(*)}, & t < 0 \\
- \sigma_{3}^{(*)} \otimes \sigma_{1}^{(A)} \otimes {\bf 1}^{(B)}, \hspace{0.6in} & t \ge 0.
\end{array}
\right. 
\end{eqnarray} 
\

In order to build the pseudo-classical paths that describe this composite system we choose an enlarged complete set of commuting observables $\{\sigma_{\rho}^{(*)}, \sigma_1^{(A)}, \sigma_{\phi}^{(B)} \}$ and expand the wavefunction (\ref{the_enlarged_state}) in the orthonormal basis of their common eigenstates: 

\begin{equation}
|\pm^{(*)}, \pm^{(A)}, \pm^{(B)}>\equiv|\pm>^{(*)} \otimes |\pm^{(A)},\pm^{(B)}>,
\end{equation}
 where $|\pm>^{(*)}$ are the polarization eigenstates of the measuring device under the operator $\sigma_{\rho}^{(*)}$ and $|\pm^{(A)},\pm^{(B)}>$ are the polarization eigenstates (\ref{E1},\ref{E2},\ref{E3},\ref{E4}) of the measured two photons system under the operators $\{\sigma_1^{(A)}, \sigma_{\phi}^{(B)} \}$ . Each of these eight paths occur with probability:

\begin{eqnarray*}
p(\pm^{(*)},\pm^{(A)},\pm^{(B)}) = \left|<\pm^{(*)},\pm^{(A)},\pm^{(B)}|{\widetilde \Psi}>\right|^2 = \left|<\pm| \downarrow >^{(*)} \right|^2 \cdot \left|<\pm,\pm|\Psi> \right|^2  = p_{(*)}(\pm) \cdot p(\pm,\pm).
\end{eqnarray*} 

We then proceed to assign to any other physical observable ${\cal O}(t)$  a polynomial operator

\begin{equation}
P_{o(t)} \left({\bf 1}, \sigma_{\rho}^{(*)}, \sigma_1^{(A)}, \sigma_{\phi}^{(B)},  \sigma_{\rho}^{(*)} \otimes \sigma_1^{(A)}, \sigma_{\rho}^{(*)} \otimes \sigma_{\phi}^{(B)},  \sigma_1^{(A)} \otimes \sigma_{\phi}^{(B)},  \sigma_{\rho}^{(*)} \otimes \sigma_1^{(A)} \otimes \sigma_{\phi}^{(B)} \right),
\end{equation} 
according to the rule:

\begin{equation}
{\cal O}(t)|{\widetilde \Psi} > = P_{o(t)} |{\widetilde \Psi} >.
\end{equation}
The perfect correlation between the polarization of photon A along the X-direction and the polarization of the device along the Z-direction as a result of the measurement (\ref{the_dynamics}) is encoded in the relationship

\begin{equation}
\sigma_3^{(*)}(t) |{\widetilde \Psi}> = - \sigma_{3}^{(*)} \otimes \sigma_{1}^{(A)} \otimes {\bf 1}^{(B)}  |{\widetilde \Psi}> =
\sigma_{1}^{(A)} |{\widetilde \Psi}> = \sigma_{1}^{(A)}(t) |{\widetilde \Psi}>, \hspace{0.6in} t \ge 0,
\end{equation}
which implies that along all paths

\begin{equation}
\left( \sigma_{3}^{(*)}  \right)_{cl}(t) = \left( \sigma_{1}^{(A)} \right)_{cl}(t) =-\left( \sigma_{1}^{(B)} \right)_{cl}(t) , \hspace{0.6in} t \ge 0.
\end{equation}
The measurement thus allows us to gain knowledge about the polarization of the photons A and B along the X-direction without actually disturbing it. Furthermore, we have seen (\ref{3stars}) that the measurement does neither disturb the polarization of photon B along the two other directions $Y,Z$. Therefore, our statistical model should properly describe the
state of photon B at emission once we have updated the probabilities of the pseudo-classical paths with the information provided by the measurement.  

Let say, for example, that the measurement returns that after the interaction $\left( \sigma_{3}^{(*)}  \right)_{cl}(t \ge 0) = +1$. Hence, $\left( \sigma_{1}^{(A)} \right)_{cl}(t \in {\bf R}) =-\left( \sigma_{1}^{(B)} \right)_{cl}(t \in {\bf R})=+1$.
The probabilities of all paths need to be updated in accordance with this information, discarding those four paths $(\pm^{(*)},-^{(A)},\pm^{(B)})$ along which  $\left( \sigma_{3}^{(*)}  \right)_{cl}(t \ge 0) = \left(\sigma_1^{(A)}\right)_{cl}=-1$ and re-evaluating according to Bayes law the probabilities to happen of the other four paths:

\begin{eqnarray*}
p(\pm^{(*)},-^{(A)},\pm^{(B)}) \rightarrow 0,
\end{eqnarray*}
\begin{eqnarray*}
p(\pm^{(*)},+^{(A)},\pm^{(B)}) \rightarrow \frac{p(\pm^{(*)},+^{(A)},\pm^{(B)})}{P},
\end{eqnarray*}
with  $P = p(+^{(*)},+^{(A)},+^{(B)}) \ + p(-^{(*)},+^{(A)},+^{(B)}) \ + \ p(+^{(*)},+^{(A)},-^{(B)}) + \ p(-^{(*)},+^{(A)},-^{(B)})$. As we are interested in the polarization of photon B along the latter four paths we integrate out the degrees of freedom of the measurement device. We are then left with two coarser paths $\pm$ occurring with probabilities,

\begin{eqnarray*}
p_{(B)}(+) \equiv p(+^{(*)},+^{(A)},+^{(B)}) + \ p(-^{(*)},+^{(A)},+^{(B)}) = \frac{1}{2} \left(1 - \cos \left(\phi\right)\right), \\
p_{(B)}(-) \equiv p(+^{(*)},+^{(A)},-^{(B)}) + \ p(-^{(*)},+^{(A)},-^{(B)}) = \frac{1}{2} \left(1 + \cos \left(\phi\right)\right).
\end{eqnarray*}
Along these two paths the polarization of photon B is given by 

\begin{eqnarray}
\label{x-x}
\left(\sigma_1^{(B)}\right)_{cl}(\pm) \ = -1,
\end{eqnarray}

\begin{eqnarray}
\label{x-y}
\left(\sigma_2^{(B)}\right)_{cl}(\pm) \ = \   \sin^{-1}\left(\phi \right) \left[\pm 1 + \cos\left(\phi\right) \right],
\end{eqnarray}

\begin{eqnarray}
\label{x-z}
\left(\sigma_3^{(B)}\right)_{cl}(\pm) & = & -i \sin^{-1}\left(\phi \right)\left[ \pm 1 + \cos\left(\phi\right) \right]. 
\end{eqnarray}
Any other observable defined on the Hilbert state of photon B can be written as a linear combination of these three observables and their values on paths can then be obtained from theirs. It is straightforward to check that this family of pseudo-classical paths (\ref{x-x},\ref{x-y},\ref{x-z}) actually describes the quantum state $|x->^{(B)}$. 

Similarly, if the measurement would have returned $\left( \sigma_{3}^{(*)}  \right)_{cl}(t \ge 0) = -1$, we would obtain (after updating according to Bayes law the probabilities of all pseudo-classical paths) that photon B was emitted in the quantum state $|x+>^{(B)}$. Hence, the notion of collapse of the wavefunction as a result of a strong measurement can be easily understood in the formalism of pseudo-classical pahs that we have developed as an update of our knowledge of the system.
\\

{\bf 7.} In the preceding sections we have shown how to build a statistical interpretation of the singlet polarization state of two entangled photons (\ref{Bell_state}) in terms of a few non-interfering coarse pseudo-classical paths with well-defined probabilities. Every physical observable of the quantum theory is given along each one of these paths a well-defined time-dependent c-value, but we have noticed above that these values are not constrained to fulfill standard algebraic relationships. Hence, it seems natural to explore under which conditions these relationships can be recovered. Such conditions could be interpreted as the onset of classicality. 

We define the {\it covariance} of any pair of physical observables ${\cal O}_1(t_1)$, ${\cal O}_2(t_2)$ along each one of the coarse paths as: 
\begin{eqnarray*}
\label{covariance}
cov_{\pm,\pm}\left[{\cal O}_1(t_1), {\cal O}_2(t_2)\right] \equiv \left({\cal O}_1(t_1) \cdot {\cal O}_2(t_2)\right)_{cl}(\pm,\pm) - \left({\cal O}_1(t_1)\right)^*_{cl}(\pm,\pm) \cdot \left( {\cal O}_2(t_2)\right)_{cl}(\pm,\pm).
\end{eqnarray*}
A path $\pi=\pm,\pm$ will appear to be a {\it classical} path \cite{david1, david2} if for any pair of physical observables the modulus of their covariance is small enough,

\begin{equation}
\left|cov_{\pi}\left[{\cal O}_1(t_1), {\cal O}_2(t_2)\right] \right| < \Delta.
\end{equation}
In particular, along such paths:

\begin{equation}
\left(\left\{{\cal O}_1(t_1) , {\cal O}_2(t_2)\right\}\right)_{cl} = \left({\cal O}_1(t_1) \cdot {\cal O}_2(t_2) + {\cal O}_2(t_2) \cdot {\cal O}_1(t_1) \right)_{cl} \simeq 2 \
{\cal R}{\it e}\left[\left({\cal O}_1(t_1)\right)^*_{cl} \cdot \left( {\cal O}_2(t_2)\right)_{cl}\right],
\end{equation}

\begin{equation}
\left(-i \left[{\cal O}_1(t_1) , {\cal O}_2(t_2)\right]\right)_{cl} = -i \left({\cal O}_1(t_1) \cdot {\cal O}_2(t_2) - {\cal O}_2(t_2) \cdot {\cal O}_1(t_1) \right)_{cl} \simeq 2 \ {\cal I}{\it m}\left[\left({\cal O}_1(t_1)\right)^*_{cl} \cdot \left( {\cal O}_2(t_2)\right)_{cl}\right]
\end{equation}
\

Therefore, the equations of motion (\ref{pseudo_equation}) that describe the time-evolution of pseudo-classical values of physical observables along these paths can be approximated as follows:

\begin{equation}
\label{pseudo_equation_classical}
\frac{d\left({\cal O}(t)\right)_{cl}}{dt}= i \left([H,{\cal O}(t)]\right)_{cl} \simeq -2 \ {\cal I}{\it m}\left[\left({\cal H}\right)^*_{cl} \cdot \left( {\cal O}(t)\right)_{cl}\right] = i \left(\left({\cal H}\right)^*_{cl} \cdot \left( {\cal O}(t)\right)_{cl} - \left( {\cal O}(t)\right)^*_{cl} \cdot \left({\cal H}\right)_{cl}\right).
\end{equation}
 \

\

{\bf 8.} In this paper we have developed a statistical interpretation of Bohm's system of two photons in their singlet polarization state. Our formalism allows to integrate the principle of locality and an extended notion of physical realism within the framework of Quantum Mechanics and, thus, it can solve the EPR paradox. \ 

We started by noticing that given the quantum wavefunction $|\Psi>$ of the photons singlet polarization state (\ref{Bell_state}) and an orthonormal basis $\{|\Psi^{(i)}_{out}>\}_{i=1,2,3,4}$ in their Hilbert space we can define a set of four non-interfering paths as follows:

\begin{itemize}
\item  The probability of each of the paths to occur is given by $p_i = \left|<\Psi^{(i)}_{out}|\Psi>\right|^2$, \ $i=1,2,3,4$.

\item Every physical observable ${\cal O}$ takes along each of these paths the time-dependent value:

\begin{equation}
\label{one_more}
\left({\cal O} \right)^{(i)}_{cl}(t) = \frac{<\Psi^{(i)}_{out}|{\cal O}(t)|\Psi>}{<\Psi^{(i)}_{out}|\Psi>}, \ i=1,2,3,4,
\end{equation}
where ${\cal O}(t) = e^{+ i H t} \ {\cal O} \ e^{-i H t}$ and $H$ is the hamiltonian of the system. These values coincide with the so-called weak values of the physical observable for determined post-selection conditions.
\end{itemize}
We call these paths pseudo-classical because the values of physical observables do not necessarily fulfill classical algebraic relationships, i.e., $\left(\{{\cal O}_1(t_1),{\cal O}_2(t_2)\}\right)^{(i)}_{cl}$ and $\left(i[{\cal O}_1(t_1),{\cal O}_2(t_2)]\right)^{(i)}_{cl}$ are not necessarily equal to $\left({\cal O}_1(t_1)\right)^{(i)*}_{cl} \cdot \left({\cal O}_2(t_2)\right)^{(i)}_{cl}+\left({\cal O}_2(t_2)\right)^{(i)*}_{cl} \cdot \left({\cal O}_1(t_1)\right)^{(i)}_{cl}$ and $i \left({\cal O}_1(t_1)\right)^{(i)*}_{cl} \cdot \left({\cal O}_2(t_2)\right)^{(i)}_{cl}-i \left({\cal O}_2(t_2)\right)^{(i)*}_{cl} \cdot \left({\cal O}_1(t_1)\right)^{(i)}_{cl}$ , respectively.
\\

This statistical model reproduces the quantum average values and two-points correlations of all physical observables:

\begin{itemize}
\item  
\begin{eqnarray*}
<\left({\cal O} \right)_{cl}(t)> \equiv \sum_{i=1,2,3,4} \ p_i \cdot \left({\cal O} \right)^{(i)}_{cl}(t) = \sum_{i=1,2,3,4} \left|<\Psi^{(i)}_{out}|\Psi>\right|^2 \cdot  \frac{<\Psi^{(i)}_{out}|{\cal O}(t)|\Psi>}{<\Psi^{(i)}_{out}|\Psi>} = \\
= \sum_{i=1,2,3,4} <\Psi|\Psi^{(i)}_{out}> \  <\Psi^{(i)}_{out}|{\cal O}(t)|\Psi> = <\Psi|{\cal O}(t)|\Psi>,
\end{eqnarray*}
\item
\begin{eqnarray*}
<\left({\cal O}_1 \right)^*_{cl}(t_1) \cdot \left({\cal O}_2 \right)_{cl}(t_2)> \equiv \sum_{i=1,2,3,4} \ p_i \cdot  \left({\cal O}_1 \right)^*_{cl}(t_1) \cdot \left({\cal O}_2 \right)_{cl}(t_2) = \hspace{1.1in} \\
 = \sum_{i=1,2,3,4} \left|<\Psi^{(i)}_{out}|\Psi>\right|^2 \cdot \left(\frac{<\Psi^{(i)}_{out}|{\cal O}_1(t_1)|\Psi>}{<\Psi^{(i)}_{out}|\Psi>}\right)^* \cdot \frac{<\Psi^{(i)}_{out}|{\cal O}_2(t_2)|\Psi>}{<\Psi^{(i)}_{out}|\Psi>}= \hspace{0.4in} \\
= \sum_{i=1,2,3,4} <\Psi|{\cal O}_1(t_1)|\Psi^{(i)}_{out}> \  <\Psi^{(i)}_{out}|{\cal O}_2(t_2)|\Psi> = <\Psi|{\cal O}_1(t_1)\cdot{\cal O}_2(t_2)|\Psi>.
\end{eqnarray*}
\end{itemize}
\

Furthermore, this statistical model is explicitly local: it can be strictly proven that the values on paths of physical properties of one of the photons do not change as a result of an interaction of the other photon with the external world or a measuring device. This feature is a direct consequence of the fact that $[{\cal O}_A, H_B]=0$, where ${\cal O}_A$ is a physical observable defined on photon A and $H_B$ is the hamiltonian that describes the interaction of photon B with the external world, which implies ${\cal O}_A(t) = e^{+ i H_B t} {\cal O}_A e^{-i H_B t} = {\cal O}_A$ and, therefore, $\left({\cal O}_A(t)\right)^{(i)}_{cl} = \left({\cal O}_A\right)^{(i)}_{cl}$ along all paths $i=1,2,3,4$.
 \\

This statistical model opens the way to solve the EPR paradox, because it describes the quantum state in terms of non-interefering paths and, furthermore, a measurement of one of the photons does not modify the predetermined values of physical observables of the other photon. Thus, after such a measurement we just need to update the probabilities of those paths that comply with its output, as we would do with a classical statistical system.
This exercise is worked out in detailed in section 6. \\ 

Nonetheless, before this formalism could be considered a solution to the EPR paradox one more issue needed to be addressed. Namely, we can actually build an infinite continous family of different statistical representations of the quantum state $|\Psi>$ through a rotation of the orthonormal linear basis $\{|\Psi^{(i)}_{out}>\}_{i=1,2,3,4}$ in its Hilbert space and the probabilities $p_i=\left|<\Psi^{(i)}_{out}|\Psi>\right|^2$ of their associated pseudo-classical paths are not invariant under this transformation. Therefore, we were led to explore how all these different representations can describe a unique common underlying reality. 

We found that the pseudo-classical paths associated to all these representations are indeed different coarse descriptions of one common finely resolved pseudo-classical statistical model, which we built step-by-step in section 5. This finely resolved statistical model, thus, solves the EPR paradox. 

The model consists of infinitely many equally probable paths distributed over the unit circle.  In order to overcome the constraints imposed by Bell's theorem the number density distribution over the circle of these finely resolved paths, as well as the actual values of physical observables along them, are defined with respect to each particular observer of either photon A or B.  We interpret this dependence as a gauge freedom.

By choosing a complete set of commuting observables $\{\sigma_1^{(A)}, \sigma_{\phi}^{(B)}\}$ to be strongly measured we define a partition of the unit circle into four coarse subspaces $(\pm,\pm)$, each one comprising all those finely resolved paths that comply with one of the four possible outputs for the specified strong measurements. The probability of each of these four coarse paths is given by the integral of the number density distribution of states over the corresponding subset of the unit circle. Moreover, the value of each physical observable on each of these coarse paths is defined by its average value over the corresponding subset of the unit circle. These four coarse paths are identical to the paths that we associated above to the linear basis  $\{|\Psi^{(i)}_{out}>\}_{i=1,2,3,4}$ defined by the eigenstates of the chosen set of commuting observables $\{\sigma_1^{(A)}, \sigma_{\phi}^{(B)}\}$. Obviously, different choices of the set of commuting observables lead to different partitions of the space of finely resolved paths, with different probabilities for the four resulting coarsely resolved paths.   \\

We think that this formalism may also become a useful tool for performing numerical simulations of quantum systems. Furthermore, our formalism may offer a new insight on the onset of classicality, without any reference to a measurement apparatus or an environment: a pseudo-classical path would appear classical if   the standard algebraic classical relationships are recovered for the values of all macroscopic physical observables. This issue will be studied in a separated paper. \\

\begin{center}
****************************** \\
\vspace{0.2in}
\end{center}

{\bf APPENDIX.} In this last section we present a brief overview of the standard descriptions of  strong and weak measurements of quantum systems. As we have done throughout the paper, we denote by $|\Psi>$ the wavefunction that describe the state of the quantum system to be measured and by ${\cal O} = \sum_{\lambda \in \Lambda} \lambda \ |\lambda> <\lambda|$ the hermitic operator with eigenvalues $\lambda \in \Lambda$ and eigenstates $\{|\lambda>\}_{\lambda \in \Lambda}$ that describe the physical observable that we intend to measure. \\

In order to perform a strong measurement of this observable we need to prepare the measuring device in a quantum eigenstate $|q=0>$ of certain operator ${\cal Q}$ and then let it interact with the measured system for a very brief time interval through a hamiltonian $H_{int} = \eta \ \delta(t) \ {\cal O}\cdot{\cal P}$, where ${\cal P}$ is the conjugate momentum of operator ${\cal Q}$ and $\eta$ is a real parameter that controls the strength of the interaction. Hence,  the entangled state of the measured system and the measuring device after their interaction is described by the wavefunction, 

\begin{eqnarray*}
e^{-i \eta \left(\sum_{\lambda \in \Lambda}\lambda |\lambda><\lambda|\right) \cdot {\cal P}}|\Psi> \otimes |q=0> = \prod_{\lambda \in \Lambda} e^{-i \eta \lambda|\lambda><\lambda| \cdot {\cal P}}|\Psi> \otimes |q=0> = \\ = \prod_{\lambda \in \Lambda} \left({\bf 1} - |\lambda><\lambda| \left({\bf 1} - e^{-i \eta \lambda {\cal P}}\right)\right)|\Psi> \otimes |q=0> = \\
= \sum_{\lambda \in \Lambda} |\lambda><\lambda| e^{-i \eta \lambda {\cal P}} |\Psi> \otimes |q=0> = \\ = 
\sum_{\lambda \in \Lambda} <\lambda|\Psi>  |\lambda> \otimes |q=\eta \lambda>.
\end{eqnarray*}
As a result of the measurement the state of the measuring device is assumed to collapse into one of the eigenstates $|q = \eta \lambda>$, $\lambda \in \Lambda$ of the observable ${\cal Q}$, leaving the measured system in the corresponding eigenstate $|\lambda>$ of the observable ${\cal O}$. Thus, the shift in the position of the pointer $\Delta = \eta \lambda$ is proportional to the output $\lambda$ of the strong measurement. \\

In order to perform a weak measurement of the same observable ${\cal O}$ we need to prepare the measuring device
in a quantum state for which the observable ${\cal Q}$ is not well defined, with average value $<{\cal Q}>=0$ and variance $<{\cal Q}^2> = \Delta q^2$. We denote this state as $|z=0>$. In addition we need to take the intensity of the interaction to be very low, $\eta \ll \Delta q$. Under these conditions the state of the system and the measuring device after their interaction is described by the wavefunction, 

\begin{eqnarray*}
e^{-i \eta {\cal O} \cdot {\cal P}}|\Psi> \otimes |q=0> \simeq
\left({\bf 1} - i \eta \ {\cal O} \cdot {\cal P}\right)|\Psi> \otimes |z=0>.
\end{eqnarray*}
\

We can now perform a strong measurement of any other observable of the system ${\cal E}$ and post-select it in the state $|\Psi_{out}>$ (an eigenstate of observable ${\cal E}$). The post-slection leaves the measuring device in the state:

\begin{eqnarray*}
<\Psi_{out}| e^{-i \eta \ {\cal O} \cdot {\cal P}}|\Psi> \otimes |z=0> \simeq <\Psi_{out}|\left({\bf 1} - i \eta \ {\cal O} \cdot {\cal P}\right)|\Psi> \otimes |z=0> = \\ = \left(<\Psi_{out}|\Psi> {\bf 1} - i \eta \ <\Psi_{out}|{\cal O}|\Psi> \cdot {\cal P}\right) \otimes |z=0> = \\ = <\Psi_{out}|\Psi> \left({\bf 1} - i \eta \frac{<\Psi_{out}|{\cal O}|\Psi>}{<\Psi_{out}|\Psi>} \cdot {\cal P}\right) |z=0> \simeq \\ \simeq <\Psi_{out}|\Psi> e^{- i \eta \frac{<\Psi_{out}|{\cal O}|\Psi>}{<\Psi_{out}|\Psi>} \cdot {\cal P}}|z=0> = \\ = <\Psi_{out}|\Psi> |z=\eta \frac{<\Psi_{out}|{\cal O}|\Psi>}{<\Psi_{out}|\Psi>}>.
\end{eqnarray*}
That is, as a result of a weak measurement and a strong post-selection the pointer shifts on average by an amount $\Delta = \eta \frac{<\Psi_{out}|{\cal O}|\Psi>}{<\Psi_{out}|\Psi>}$, proportional to the weak value of the measured observable ${\cal O}$. As this shift maybe much smaller than the actual width of the pointer wavefunction $\Delta q$, a weak measurement must be performed on many identical systems in order to get an accurate estimation.

\end{document}